\documentclass[aps, prd,12pt, nofootinbib, floatfix, showpacs, superscriptaddress]{revtex4-2} 
\usepackage{amsmath, amssymb, mathtools, bm, bbm, tabularx, subfigure, graphicx,placeins}
\allowdisplaybreaks[0]
\usepackage[dvipsnames]{xcolor}
\usepackage{hyperref}
\usepackage{youngtab}\Yboxdim 6.5pt\Ylinethick 0.4pt
\usepackage{natbib}
\graphicspath{{Figures/}}

\begin{document}

\title{The hidden sector variations in the ${\cal N}=1$ supersymmetric three-family Pati-Salam models  from intersecting D6-branes}

\author{Adeel Mansha}
\email{adeelmansha@zjnu.edu.cn}
\affiliation{Department of Physics, Zhejiang Normal University, Jinhua 321004, P. R. China}
\affiliation{Zhejiang Institute of Photoelectronics, Zhejiang Normal University, Jinhua, Zhejiang 321004, China}
\affiliation{Zhejiang Institute for Advanced Light
Source, Zhejiang Normal University, Jinhua, Zhejiang 321004, China}

\author{Tianjun Li}
\email{tli@itp.ac.cn}
\affiliation{CAS Key Laboratory of Theoretical Physics, Institute of Theoretical Physics, Chinese Academy of Sciences, Beijing 100190, P. R. China}
\affiliation{School of Physical Sciences, University of Chinese Academy of Sciences, Beijing, P. R. China}

\author{Lina Wu}
\thanks{Corresponding author}
\email{wulina@xatu.edu.cn}
\affiliation{School of Sciences, Xi'an Technological University, Xi'an 710021, P. R. China}

\date{\today}

\begin{abstract}

It is well-known that three-family supersymmetric Pati-Salam models from intersecting D6-branes, where either one or both of the ${\rm U}(2)$ gauge factors are replaced by a ${\rm USp}(2)$ group,  are quite scarce. In order to construct all such kind of models with generic additional gauge symmetries, we fix the observable sectors and study all the possible hidden sectors. Thus, we are able to completely determine all types of such kind of the inequivalent models on a $\mathbbm{T}^6/(\mathbbm{Z}_2\times \mathbbm{Z}_2)$ orientifold from IIA string theory. We find the gauge coupling relations to be highly sensitive to the variations in the hidden sector. One of the models exhibits the gauge coupling unification for a particular solution at the string scale. In addition, we perform scan on the hidden sector variations for the models presented in \href{https://arxiv.org/abs/2112.09632}{arXiv:2112.09632}, whose gauge coupling relations are still preserved. Thus, we fix the gap in the previous study and complete the model building for all the inequivalent models.

\end{abstract}

\maketitle

\section{Introduction}\label{sec:Intro}

Intersecting D6-branes in type IIA string theory have been extremely useful to geometrically understand the standard model (SM) from a certain Calabi-Yau compactification of extra dimensions. The gauge structures, the chiral spectra and the various couplings arise from the D-brane configurations. For example, the four-dimensional gauge couplings depend on the volume of the cycles wrapped by the D-branes and the gravitational coupling is determined by their total internal volume. Similarly, the cubic Yukawa couplings depend exponentially on the triangular areas of open worldsheet intersections. The general flavor structure and selection rules for intersecting D-brane models has been investigated in \cite{Chamoun:2003pf, Higaki:2005ie}.

The SM fermions belong to the chiral representations of the gauge group ${\rm SU}(3)_C\times {\rm SU}(2)_W \times {\rm U}(1)_Y$ such that all gauge anomalies are canceled. Simply placing parallel D-branes in flat space does not yield chiral fermions. Instead, to realize the chiral fermions we need to either place D-branes on orbifold singularities \cite{Aldazabal:2000sa} or else consider intersecting D-branes on generalized orbifolds called orientifolds \cite{Shiu:1998pa, Cvetic:2001tj}. In orientifolds, both the discrete internal symmetries of the world-sheet theory and the products of internal symmetries with world-sheet parity reversal become gauged.

The ${\cal N}=1$ supersymmetric three-family Pati-Salam models in the IIA string theory  on $\mathbbm{T}^6/(\mathbbm{Z}_2\times \mathbbm{Z}_2)$ orientifold with intersecting D6-branes have been constructed in ref.~\cite{Cvetic:2004ui}, where the wrapping number is up to 3. Recently, utilizing machine learning algorithm in ref.~\cite{Li:2019nvi} several new models with the wrapping number up to 5 were consrtructed. The phenomenology of models up to the wrapping number of 3 was first studied in ref.~\cite{Chen:2007zu} while the phenomenology of a newly found model with a wrapping number 5 was explored in \cite{Sabir:2022hko}. In ref.~\cite{He:2021gug,He:2021kbj}, a mathematical search algorithm was used to possibly obtain the complete landscape of supersymmetric Pati-Salam models, where again the highest wrapping number turned out to be 5 as already indicated from the random search in \cite{Li:2019nvi}.
However, an interesting variation of the Pati-Salam gauge symmetry $U(4)_C\times U(2)_L \times U(2)_R$
that was not taken into account in previous searches \cite{Cvetic:2004ui, Li:2019nvi, He:2021gug,He:2021kbj} is the ${\rm U}(4)_C \times {\rm USp}(2)_L \times {\rm U}(2)_R$ or ${\rm U}(4)_C \times {\rm USp}(2)_L \times {\rm USp}(2)_R$ where either one or both of the usual $U(2)$s are replaced by $USp(2)$ groups. The construction of $USp(2)$ groups can be readily achieved by taking the corresponding stacks of D6-branes parallel to any of the O6-planes or their $\mathbbm{Z}_2$ images as exemplified in refs.~\cite{Cvetic:2004nk, Mansha:2022pnd}. The choice of $USp(2)$ is simpler to deal with since there is no associated global anomalous $U(1)$ group which is generic in the unitary groups.

In this paper, we fix the visible sector stacks $a$, $b$ and $c$ of the supersymmetric Pati-Salam models with the ${\rm USp(2)}_{L,R}$ group and vary the hidden sector to search for all possible gauge group factors consistent with the four-dimensional ${\cal N}=1$ supersymmetry conditions and tadpole cancellation conditions.
This systematic search further pins down the allowed number of gauge group factors for the supersymmetric Pati-Salam Models from intersecting D6-branes \cite{Chen:2007ms}. Taking the two new models as examples, we show the string-scale gauge coupling relations can be realized through two-loop level renormalization group equation (RGE) running by introducing vector-like particles. In \cite{Li:2022cqk}, the additional vector-like matters are introduced to push GUT scale unification up to string scale for the $\mathcal{N}=1$ supersymmetric Pati-Salam models. The unification of three independent gauge interactions into a single one has been extensively studied previously from various physical perspectives \cite{Bachas:1995yt, Lopez:1996gd, Blumenhagen:2003jy, Barger:2005qy, Jiang:2006hf, Barger:2007qb, Jiang:2008xrg, Jiang:2009za,Kokorelis:2016ckp, Chen:2017rpn, Chen:2018ucf,Aranda:2020fkj,Maru:2022mbi}. Moreover, to complete the model building, the machine learning is performed  to scan on the hidden sector variations for the models listed in Ref.~\cite{He:2021gug} with the fixed visible sector.

The plan of the paper is as follows. In section~\ref{sec:orientifold}, we briefly review the model building rules for $\mathcal{N}=1$ supersymmetric Pati-Salam models with the gauge group ${\rm SU}(4)_C \times {\rm USp}(2)_L \times {\rm USp}(2)_R$ on a $\mathbbm{T}^6/(\mathbbm{Z}_2\times \mathbbm{Z}_2)$ orientifold. In section~\ref{sec:Pheno}, we discuss the salient phenomenological features of the newly engineered models under the variation of the hidden sector gauge group and list the perturbative particle spectra of the models. In section~\ref{sec:RGE}, we investigate the RGE running for the gauge couplings at two-loop level and obtain the gauge coupling relations at string-scale. We also perform machine learning in section~\ref{sec:ml} to scan on the hidden sector variations for the models. Finally, we conclude in section~\ref{sec:conclusion}.

\section{Pati-Salam model building from $\mathbbm{T}^6/(\mathbbm{Z}_2\times \mathbbm{Z}_2)$ orientifold} \label{sec:orientifold}

The basic rules to construct the supersymmetric Pati-Salam models from Type IIA $\mathbbm{T}^6/(\mathbbm{Z}_2\times \mathbbm{Z}_2)$ orientfolds with intersecting D6-branes have been discussed in ref.~\cite{Cvetic:2004ui}. We construct the Pati-Salam models with at least one symplatic group by following the conventions in similar fashion as discussed in ref. \cite{Mansha:2022pnd}. $\mathbbm{T}^6$ is a factorisable six-dimensional torus i.e $\mathbbm{T}^6 = \mathbbm{T}^2 \times \mathbbm{T}^2 \times \mathbbm{T}^2$ in the  $\mathbbm{T}^6/(\mathbbm{Z}_2\times \mathbbm{Z}_2)$ orientifolds, and  $\theta$ and $\omega$ are the generators of orbifold group $(\mathbbm{Z}_2\times \mathbbm{Z}_2)$ which act on the complex coordinates $z_i$ as,
\begin{align}
	\theta: \quad (z_1,z_2,z_3) &\to (-z_1,-z_2,z_3), \nonumber \\
	\omega: \quad (z_1,z_2,z_3) &\to (z_1,-z_2,-z_3). \label{orbifold}
\end{align}
Orientifold projection is implemented by gauging the $\Omega R$ symmetry, where $\Omega$ is world-sheet parity, and $R$ acts on the complex coordinates as,
\begin{align}
	R: \quad (z_1,z_2,z_3) &\to (\Bar{z_1},\Bar{z_2},\Bar{z_3}).
\end{align}
This leads to the appearance of four different kinds of orientifold 6-planes (O6-planes) corresponding to $\Omega R$, $\Omega R\theta$, $\Omega R\omega$, and $\Omega R\theta\omega$ respectively. There are two different kinds of complex structures which are consistent with orientifold projection $i.e$ rectangular, and tilted. In order to construct three families of the SM fermions, it is required to have at least one two-torus to be tilted. So, in our model building setup, the last tow-torus is tilted $i.e$ $\beta_3=1$, and, $\beta_{1,2}=0$. So, the homology three-cycles wrapped by stack $a$ of $N_a$ D6-branes with the cycle $(n_a^i,l_a^i)$ and their $\Omega R$ images ${a'}$ take the form as follows,
\begin{align}
	[\Pi_a ]&=\prod_{i=1}^{3}\left(n_{a}^{i}[a_i]+2^{-\beta_i}l_{a}^{i}[b_i]\right), \nonumber \\
	[\Pi_{a'}] &=\prod_{i=1}^{3}\left(n_{a}^{i}[a_i]-2^{-\beta_i}l_{a}^{i}[b_i]\right),
\end{align}
where $\beta_i=0$ or $\beta_i=1$ for the rectangular or tilted $i^{\rm th}$ two-torus, respectively.
And the homology three-cycles wrapped by the four O6-planes take the form,
\begin{alignat}{2}
	\Omega R :            &\quad&         [\Pi_{\Omega R}] &= 2^3 [a_1]\times[a_2]\times[a_3],  \nonumber\\
	\Omega R\omega :      &&        [\Pi_{\Omega R\omega}] &=-2^{3-\beta_2-\beta_3}[a_1]\times[b_2]\times[b_3],  \nonumber\\
	\Omega R\theta\omega : && [\Pi_{\Omega R\theta\omega}] &=-2^{3-\beta_1-\beta_3}[b_1]\times[a_2]\times[b_3], \nonumber\\
	\Omega R\theta :      &&       [\Pi_{\Omega R \theta}] &=-2^{3-\beta_1-\beta_2}[b_1]\times[b_2]\times[a_3]. \label{orienticycles}
\end{alignat}
Therefore, the intersection numbers can be written as,
\begin{align}
	I_{ab} &=[\Pi_a][\Pi_b] =2^{-k}\prod_{i=1}^3(n_a^il_b^i-n_b^il_a^i),\nonumber\\
	I_{ab'}&=[\Pi_a]\left[\Pi_{b'}\right] =-2^{-k}\prod_{i=1}^3(n_{a}^il_b^i+n_b^il_a^i),\nonumber\\
	I_{aa'}&=[\Pi_a]\left[\Pi_{a'}\right] =-2^{3-k}\prod_{i=1}^3(n_a^il_a^i),\nonumber\\
	I_{aO6}&=[\Pi_a][\Pi_{O6}] =2^{3-k}(-l_a^1l_a^2l_a^3+l_a^1n_a^2n_a^3+n_a^1l_a^2n_a^3+n_a^1n_a^2l_a^3),\label{intersections}
\end{align}
where $k=\sum_{i=1}^3\beta_i$ and $[\Pi_{O6}]=[\Pi_{\Omega R}]+[\Pi_{\Omega R\omega}]+[\Pi_{\Omega R\theta\omega}]+[\Pi_{\Omega R\theta}]$.

\subsection{The RR tadpole cancellation and supersymmetry conditions}\label{subsec:constraints}

Since, the sources of RR fields, D6-branes, and O6-planes need to satisfy the Gauss's law $i.e$ the total RR charges must
vanish as the RR field flux lines are conserved \cite{Gimon:1996rq},
\begin{eqnarray}\label{RRtadpole}
	\sum_a N_a [\Pi_a]+\sum_a N_a \left[\Pi_{a'}\right]-4[\Pi_{O6}]=0,
\end{eqnarray}
where the contribution of last term comes from the O6-planes, which have $-4$ RR charges in D6-brane charge units. The ${\rm SU}(N_a)^3$ cubic non-Abelian anomaly is cancelled by RR tadpole cancellation conditions, while U(1) mixed gauge and gravitational anomaly or $[{\rm SU}(N_a)]^2 {\rm U}(1)$ gauge anomaly can be cancelled with the Green-Schwarz mechanism~\cite{Green:1984sg}.

Let us simplify the notations by defining the following products of wrapping numbers,
\begin{alignat}{4}
	A_a &\equiv -n_a^1n_a^2n_a^3, &\quad B_a &\equiv n_a^1l_a^2l_a^3, &\quad C_a &\equiv l_a^1n_a^2l_a^3, \quad & D_a &\equiv l_a^1l_a^2n_a^3, \nonumber\\
	\tilde{A}_a &\equiv -l_a^1l_a^2l_a^3, & \tilde{B}_a &\equiv l_a^1n_a^2n_a^3, & \tilde{C}_a &\equiv n_a^1l_a^2n_a^3, & \tilde{D}_a &\equiv n_a^1n_a^2l_a^3.\,\label{variables}
\end{alignat}
In order to cancel the RR tadpoles, the contribution from an arbitrary number of D6-branes wrapped along the orientifold planes, the so called “filler branes”, can also be added which trivially satisfy the four-dimensional ${\cal N}=1$ supersymmetry conditions. Thus, the tadpole conditions take the form,
\begin{align}
	-2^k N^{(1)}+\sum_a N_a A_a &= -2^k N^{(2)}+\sum_a N_a B_a = \nonumber\\
	-2^k N^{(3)}+\sum_a N_a C_a &= -2^k N^{(4)}+\sum_a N_a D_a = -16,\,
\end{align}
where $2 N^{(i)}$ denotes the number of filler branes wrapped along the $i^{\rm th}$ O6-plane as shown in table~\ref{orientifold}.

\begin{table}[htb]
	\renewcommand{\arraystretch}{1.3}
	\caption{The wrapping numbers for four O6-planes.} \label{orientifold}
	\begin{center}
		$\begin{array}{|c|c|c|}
			\hline
			\text{Orientifold Action} & \text{O6-Plane} & (n^1,l^1)\times (n^2,l^2)\times (n^3,l^3)\\
			\hline\hline
			\Omega R& 1 & (2^{\beta_1},0)\times (2^{\beta_2},0)\times (2^{\beta_3},0) \\
			\hline
			\Omega R\omega& 2& (2^{\beta_1},0)\times (0,-2^{\beta_2})\times (0,2^{\beta_3}) \\
			\hline
			\Omega R\theta\omega& 3 & (0,-2^{\beta_1})\times (2^{\beta_2},0)\times (0,2^{\beta_3}) \\
			\hline
			\Omega R\theta& 4 & (0,-2^{\beta_1})\times (0,2^{\beta_2})\times (2^{\beta_3},0) \\
			\hline
		\end{array}$
	\end{center}
\end{table}

If the rotation angle of any D6-brane with respect to the orientifold-plane is an element of ${\rm SU}(3)$, the 4-dimensional ${\cal N}=1$ supersymmetry (SUSY) can be preserved after compactification from ten-dimensions  or SUSY condition take the form,
\begin{equation}
	\theta^a_1 + \theta^a_2 + \theta^a_3 = 0 \mod 2\pi ,
\end{equation}
with $\theta^a_j = \arctan (2^{- \beta_j} \chi_j l^a_j/n^a_j)$. $\theta_i$ is the angle between the $D6$-brane and orientifold-plane in the $i^{\rm th}$ 2-torus and $\chi_i=R^2_i/R^1_i$ are the complex structure moduli for the $i^{\rm th}$ 2-torus. The SUSY conditions can also be written as,
\begin{eqnarray}
	x_A\tilde{A}_a+x_B\tilde{B}_a+x_C\tilde{C}_a+x_D\tilde{D}_a=0,\nonumber\\
	\frac{A_a}{x_A}+\frac{B_a}{x_B}+\frac{C_a}{x_C}+\frac{D_a}{x_D} < 0, \label{susyconditions}
\end{eqnarray}
where $x_A=\lambda,\; x_B=2^{\beta_2+\beta_3}\cdot\lambda /\chi_2\chi_3,\; x_C=2^{\beta_1+\beta_3}\cdot\lambda /\chi_1\chi_3,\; x_D=2^{\beta_1+\beta_2}\cdot\lambda /\chi_1\chi_2$.

Orientifolds also have discrete D-brane RR charges classified by the $\mathbbm{Z}_2$ K-theory groups~\cite{Witten:1998cd, Cascales:2003zp, Marchesano:2004yq, Marchesano:2004xz}, which imply~\cite{Uranga:2000xp},
\begin{eqnarray}
	\sum_a \tilde{A}_a  = \sum_a  N_a  \tilde{B}_a = \sum_a  N_a  \tilde{C}_a = \sum_a  N_a \tilde{D}_a = 0 \mod 4 \label{K-charges}~.~\,
\end{eqnarray}
In Pati-Salam models, we can avoid the nonvanishing torsion charges by taking an even number of D-branes, {\it i.e.}, $N_a \in 2 \mathbbm{Z}$.

\begin{table}[htb]
	\renewcommand{\arraystretch}{1.3}\centering
	\caption{General spectrum for intersecting D6-branes at generic angles, where ${\cal M}$ is the multiplicity, and  $a_{\protect\yng(2)}$ and $a_{\protect\yng(1,1)}$ denote respectively the symmetric and antisymmetric representations of ${\rm U}(N_a/2)$. Positive intersection numbers in our convention refer to the left-handed chiral supermultiplets.\\}
	$\begin{array}{|c|c|}
		\hline
		\text{\bf Sector} & {\bf Representation}  \\
		\hline\hline
		aa                & {\rm U}(N_a/2) \text{ vector multiplet}  \\
		& \text{3 adjoint chiral multiplets}  \\
		\hline ab+ba      & {\cal M}(\frac{N_a}{2}, \frac{\overline{N_b}}{2})= I_{ab}(\yng(1)_{a},\overline{\yng(1)}_{b}) \\
		\hline ab'+b'a    & {\cal M}(\frac{N_a}{2}, \frac{N_b}{2})=I_{ab'}(\yng(1)_{a},\yng(1)_{b}) \\
		\hline aa'+a'a    & {\cal M} (a_{\yng(2)})= \frac{1}{2} (I_{aa'} -  \frac{1}{2} I_{aO6}) \\
		& {\cal M} (a_{\yng(1,1)_{}})= \frac{1}{2} (I_{aa'} + \frac{1}{2} I_{aO6})  \\
		\hline
	\end{array}$
	\label{tab:spectrum}
\end{table}

The general particle representations for intersecting D6-branes models at angles are shown in table \ref{tab:spectrum}. Following the convention of \cite{Cvetic:2004ui} the $N$ number of D6-brane stacks corresponds to U($N/2$) and USp($N$) respectively. A positive intersection number in our convention refers to the left-chiral supermultiplet.

The effective gauge symmetry of supersymmetric Pati-Salam models can be further broken down to the standard model gauge symmetry via brane splitting. The relevant details and the formulae of computing gauge kinetic functions and the gauge coupling relations are given in \cite{Cvetic:2004nk, Sabir:2022hko, Mansha:2022pnd}.

\section{Pati-Salam models under the variation of the hidden sector}\label{sec:Pheno}

One of the main objectives of our paper is to emphasize that a complete search of such three family models should not only focus on the observable sector but also take into account the variations under the hidden sector. Because, it is possible to find inequivalent models with the same observable sector while a completely different hidden sector of the D6-brane stacks and O6-planes. In the following, we discuss the models with one USp(2) group and the models with two USp(2) groups, and the models without symplectic gauge group separately.

Three family supersymmetric Pati-Salam models from symplectic groups where either one or both of the SU(2) gauge factor is replaced by USp(2) groups are of special interest because unlike the U(2) stacks the USp brane stacks are parallel to either of the O6-planes. This fact explains the relative scarcity of such models as compared to the usual models strictly arising from the unitary gauge group stacks.

\subsection{Pati-Salam models with the gauge group ${\rm U}(4)_C \times {\rm U}(2)_L \times {\rm U}(2)_R$}

Pati-Salam models with the gauge group ${\rm U}(4)_C \times {\rm U}(2)_L \times {\rm U}(2)_R$ have been recently discussed in ref.~\cite{He:2021gug,He:2021kbj}. It turns out that there are only 33 independent models with different gauge coupling relations. We fix the visible sector, and perform scan on the hidden sector which gives rise to the inequivalent models, and thus, fix the gap in their study. In this way, we clean up the search for Pati-Salam models with gauge group ${\rm U}(4)_C \times {\rm U}(2)_L \times {\rm U}(2)_R$. Since, the visible sector is fixed, the gauge coupling relations will remain same. These new models are listed as Model~\hyperref[model_1A]{1A}, Model~\hyperref[model_2A]{2A}, and Model~\hyperref[model_3A]{3A}, while their detailed particle spectra are given in Table~\ref{spectrum_model1A}, Table~\ref{spectrum_model2A}, and Table~\ref{spectrum_model3A} respectively.
	\begin{table}[htb]
		\centering \footnotesize
		\caption{Model~\hyperref[model_1A]{1A} represents the hidden sector variation of model 14 (T-dual) in ref.~\cite{He:2021gug}. D6-brane configurations and intersection numbers of Model~\hyperref[model_1A]{1A}, and its MSSM gauge coupling relation is $g^2_a=g^2_b=\frac{1}{3}g^2_c=\frac{5}{11}(\frac{5}{3}g^2_Y)=\frac{16\pi}{5\sqrt{3} }e^{\phi ^4}$.\\} \label{model_1A}
		$\begin{array}{|c||c|c| |r|r| r|  r |r| r |r| r|r|r| r|r|r|}
			\hline
			\rm{Model~\hyperref[model_1A]{1A}}  & \multicolumn{14}{c|}{{\rm U}(4)\times {\rm U}(2)_L \times {\rm U}(2)_R \times {\rm U}(2)\times {\rm USp}(2)}\\
			\hline \hline \rm{stack} & N & (n^1,l^1)\times (n^2,l^2)\times (n^3,l^3) & n_{\yng(2)} & n_{\yng(1,1)_{}} & b & b' & c  & c' & d & d' & 1 & 2 & 3 & 4 \\
			\hline
			a     & 8     & (0, -1)\times (1, 1)\times (1, 1) & 0     & 0         & 3     &   0   & -3    &   0   & 0           & 0     & 0     & 0     & 0     & 0 \\
			b     & 4     & (3, 1)\times (1, 0)\times (1, -1) & 2     & -2        & \text{-}    &  \text{-} & 0     &  4      & -3          & 0    & 0     & 0     & 0     & -3 \\
			c     & 4     & (1, 0)\times (1, 4)\times (1, -1) & -3     & 3       & \text{-}    & \text{-}    &  \text{-}    & \text{-}    & 3          & 0   & 0     & 0     & 0     & -1 \\
			d     & 4     & (0, 1)\times (-1, -1)\times (1, 1) & 0    & 0      & \text{-}    & \text{-}    & \text{-}    & \text{-}    & \text{-}    & \text{-}    & 0    & 0  & 0   & 0 \\
			\hline
			4     & 2     & (0, -1)\times (0, 1)\times (2, 0) &  \multicolumn{12}{c|}{\beta^g_d=0, \quad\beta^g_4=-2,   } \\
			&       &                                   &   \multicolumn{12}{c|}{x_A=x_B=12x_C=3x_D=1} \\
			\hline
		\end{array}$
	\end{table}
 
	\begin{table}[htb]
		\centering \footnotesize
		\caption{Model~\hyperref[model_2A]{2A} represents the hidden sector variation of model 44 in ref.~\cite{He:2021gug}. D6-brane configurations and intersection numbers of Model~\hyperref[model_2A]{2A}, and its MSSM gauge coupling relation is $g^2_a=g^2_b=g^2_c=(\frac{5}{3}g^2_Y)=4 \sqrt{\frac{2}{3}} \pi  e^{\phi ^4}     $.\\} \label{model_2A}
		$\begin{array}{|c||c|c| |r|r| r|  r |r| r |r| r|r|r| r|r|r|}
			\hline
			\rm{Model~\hyperref[model_2A]{2A}}  & \multicolumn{14}{c|}{{\rm U}(4)\times {\rm U}(2)_L \times {\rm U}(2)_R \times {\rm U}(2)\times {\rm USp}(2)\times {\rm USp}(2)}\\
			\hline \hline \rm{stack} & N & (n^1,l^1)\times (n^2,l^2)\times (n^3,l^3) & n_{\yng(2)} & n_{\yng(1,1)_{}} & b & b' & c  & c' & d & d' & 1 & 2 & 3 & 4 \\
			\hline
			a     & 8     & (1, 0)\times (1, -1)\times (1, 1) & 0     & 0         & 0     &   3   & 0    &   -3   & 0           & 0     & 0     & 0     & -1     & 1 \\
			b     & 4     & (-1, -3)\times (0, -1)\times (-1, -1) & -2     & 2        & \text{-}    &  \text{-} & 0     &  0      & 3          & 0    & 0     & 0     & -1     & 0 \\
			c     & 4     & (1, -3)\times (-1, 0)\times (-1, -1) & 2     & -2       & \text{-}    & \text{-}    &  \text{-}    & \text{-}    & -3          & 0   & 0     & 0     & 0     & 1 \\
			d     & 4     & (1, 0)\times (1, 1)\times (1, -1) & 0    & 0      & \text{-}    & \text{-}    & \text{-}    & \text{-}    & \text{-}    & \text{-}    & 0    & 0  & 1   & -1 \\
			\hline
			3     & 2     & (0, -1)\times (1, 0)\times (0, 2) &  \multicolumn{12}{c|}{\beta^g_d=0, \quad\beta^g_3=-2, \quad\beta^g_4=-2,   } \\
			4 &    2   &    (0, -1)\times (0, 1)\times (2, 0)                            &   \multicolumn{12}{c|}{x_A=x_B=3x_C=3x_D=1} \\
			\hline
		\end{array}$
	\end{table}

	\begin{table}[htb]
		\centering \footnotesize
		\caption{Model~\hyperref[model_3A]{3A} represents the hidden sector variation of model 29 (T-dual) in ref.~\cite{He:2021gug}. D6-brane configurations and intersection numbers of Model~\hyperref[model_3A]{3A}, and its MSSM gauge coupling relation is $g^2_a=\frac{13}{5}g^2_b=3g^2_c=\frac{5}{3}(\frac{5}{3}g^2_Y)=\frac{16\sqrt{3}\pi}{5 }e^{\phi ^4}$.\\} \label{model_3A}
		$\begin{array}{|c||c|c| |r|r| r|  r |r| r |r| r|r|r| r|r|r|}
			\hline
			\rm{Model~\hyperref[model_3A]{3A}}  & \multicolumn{14}{c|}{{\rm U}(4)\times {\rm U}(2)_L \times {\rm U}(2)_R \times {\rm U}(2)\times {\rm USp}(2)}\\
			\hline \hline \rm{stack} & N & (n^1,l^1)\times (n^2,l^2)\times (n^3,l^3) & n_{\yng(2)} & n_{\yng(1,1)_{}} & b & b' & c  & c' & d & d' & 1 & 2 & 3 & 4 \\
			\hline
			a     & 8     & (0, -1)\times (-1, -1)\times (-1, -1) & 0     & 0         & 1     &   2   & 0    &   -3   & 0           & 0     & 0     & 0     & 0     & 0 \\
			b     & 4     & (1, -1)\times (-1, 0)\times (-3, -1) & 2     & -2        & \text{-}    &  \text{-} & -4     &  -8      & 2          & -1    & 0     & 0     & 0     & 1 \\
			c     & 4     & (1, 0)\times (-1, 4)\times (-1, -1) & 3     & -3       & \text{-}    & \text{-}    &  \text{-}    & \text{-}    & -3          & 0   & 0     & 0     & 0     & 1 \\
			d     & 4     & (0, -1)\times (-1, 1)\times (1, -1) & 0    & 0      & \text{-}    & \text{-}    & \text{-}    & \text{-}    & \text{-}    & \text{-}    & 0    & 0  & 0   & 0 \\
			\hline
			    4 &   2   & (0, -1)\times (0, 1)\times (2, 0) &  \multicolumn{12}{c|}{\beta^g_d=-1,  \quad\beta^g_4=-4,   } \\
			 &       &                                &   \multicolumn{12}{c|}{x_A=x_B=\frac{4}{3}x_C=\frac{1}{3}x_D=1} \\
			\hline
		\end{array}$
	\end{table}

\begin{table}
[htb] \footnotesize
\renewcommand{\arraystretch}{1.0}
\caption{The chiral spectrum in the open string sector of Model~\hyperref[model_1A]{1A}} \label{spectrum_model1A}
\begin{center}
\begin{tabular}{|c||c||c|c|c||c|c|c|}\hline
Model~\hyperref[model_1A]{1A} & $SU(4)\times SU(2)_L\times SU(2)_R\times SU(2)_d \times USp(2)$
& $Q_4$ & $Q_{2L}$ & $Q_{2R}$ & $Q_{em}$ & $B-L$ & Field \\
\hline\hline
$ab$ & $3 \times (4,\overline{2},1,1,1)$ & 1 & -1 & 0  & $-\frac 13,\; \frac 23,\;-1,\; 0$ & $\frac 13,\;-1$ & $Q_L, L_L$\\
$ac$ & $3\times (\overline{4},1,2,1,1)$ & -1 & 0 & $1$   & $\frac 13,\; -\frac 23,\;1,\; 0$ & $-\frac 13,\;1$ & $Q_R, L_R$\\
$bc'$ & $4\times (4,\overline{2},\overline{2},1,1)$ & 0 & -1 & $-1$   & $-1,\; 0,\;0,\; 1$ & $0$ & $H'$\\
$bd$ & $3\times(1,\overline{2},1,2,1)$ & 0 & -1 & 0   & $\pm \frac 12$ & 0 & \\
$b4$ & $3\times(1,\overline{2},1,1,2)$ & 0 & -1 & 0   & $\mp \frac 12$ & 0 & \\
$cd$ & $3\times(1,1,2,\overline{2},1)$ & 0 & 0 & 1   & $\pm \frac 12$ & 0 & \\
$c4$ & $1\times(1,1,\overline{2},1,2)$ & 0 & 0 & -1   & $\pm \frac 12$ & 0 & \\
$b_{\yng(2)}$ & $2\times(1,3,1,1,1)$ & 0 & 2 & 0   & $0,\pm 1$ & 0 & \\
$b_{\overline{\yng(1,1)}}$ & $2\times(1,\overline{1},1,1,1)$ & 0 & -2 & 0   & 0 & 0 & \\
$c_{\overline{\yng(2)}}$ & $3\times(1,1,\overline{3},1,1)$ & 0 & 0 & -3   & $0,\pm 1$ & 0 & \\
$c_{\yng(1,1)}$ & $3\times(1,1,1,1,1)$ & 0 & 0 & 3   & 0 & 0 & \\
\hline
\end{tabular}
\end{center}
\end{table}

\begin{table}
[htb] \footnotesize
\renewcommand{\arraystretch}{1.0}
\caption{The chiral spectrum in the open string sector of Model~\hyperref[model_2A]{2A}} \label{spectrum_model2A}
\begin{center}
\begin{tabular}{|c||c||c|c|c||c|c|c|}\hline
Model~\hyperref[model_2A]{2A} & $SU(4)\times SU(2)_L\times SU(2)_R\times SU(2)_d \times USp(2)^2$
& $Q_4$ & $Q_{2L}$ & $Q_{2R}$ & $Q_{em}$ & $B-L$ & Field \\
\hline\hline
$ab'$ & $3 \times (4,2,1,1,1,1)$ & 1 & 1 & 0  & $-\frac 13,\; \frac 23,\;-1,\; 0$ & $\frac 13,\;-1$ & $Q_L, L_L$\\
$ac'$ & $3\times (\overline{4},1,\overline{2},1,1,1)$ & -1 & 0 & $-1$   & $\frac 13,\; -\frac 23,\;1,\; 0$ & $-\frac 13,\;1$ & $Q_R, L_R$\\
$a3$ & $1\times (\overline{4},1,1,1,2,1)$ & $-1$ & 0 & 0 & $-\frac 16,\;\frac 12$ & $-\frac 13,\;1$ & \\
$a4$ & $1\times (4,1,1,1,1,2)$ & 1 & 0 & 0   & $\frac 16,\;-\frac 12$ & $\frac 13,\;-1$ & \\
$bd$ & $3\times(1,2,1,\overline{2},1,1)$ & 0 & 1 & 0   & $\pm \frac 12$ & 0 & \\
$b3$ & $1\times(1,\overline{2},1,1,2,1)$ & 0 & -1 & 0   & $\mp \frac 12$ & 0 & \\
$cd$ & $3\times(1,1,\overline{2},2,1,1)$ & 0 & 0 & -1   & $\pm \frac 12$ & 0 & \\
$c4$ & $1\times(1,1,2,1,1,2)$ & 0 & 0 & 1   & $\pm \frac 12$ & 0 & \\
$d3$ & $1\times(1,1,1,2,2,1)$ & 0 & 0 & 0   & $\pm \frac 12$ & 0 & \\
$d4$ & $1\times(1,1,1,\overline{2},1,2)$ & 0 & 0 & 0   & $\pm \frac 12$ & 0 & \\
$b_{\overline{\yng(2)}}$ & $2\times(1,\overline{3},1,1,1,1)$ & 0 & -2 & 0   & $0,\pm 1$ & 0 & \\
$b_{\yng(1,1)}$ & $2\times(1,1,1,1,1,1)$ & 0 & 2 & 0   & 0 & 0 & \\
$c_{\yng(2)}$ & $2\times(1,1,3,1,1,1)$ & 0 & 0 & 2   & $0,\pm 1$ & 0 & \\
$c_{\overline{\yng(1,1)}}$ & $2\times(1,\overline{1},1,1,1,1)$ & 0 & 0 & -2   & 0 & 0 & \\
\hline
\end{tabular}
\end{center}
\end{table}

\begin{table}
[htb] \footnotesize
\renewcommand{\arraystretch}{1.0}
\caption{The chiral spectrum in the open string sector of Model~\hyperref[model_3A]{3A}} \label{spectrum_model3A}
\begin{center}
\begin{tabular}{|c||c||c|c|c||c|c|c|}\hline
Model~\hyperref[model_3A]{3A} & $SU(4)\times SU(2)_L\times SU(2)_R\times SU(2)_d \times USp(2)$
& $Q_4$ & $Q_{2L}$ & $Q_{2R}$ & $Q_{em}$ & $B-L$ & Field \\
\hline\hline
$ab$ & $1 \times (4,\overline{2},1,1,1)$ & 1 & -1 & 0  & $-\frac 13,\; \frac 23,\;-1,\; 0$ & $\frac 13,\;-1$ & $Q_L, L_L$\\
$ab'$ & $2 \times (\overline{4},\overline{2},1,1,1)$ & -1 & -1 & 0  & $-\frac 13,\; \frac 23,\;-1,\; 0$ & $\frac 13,\;-1$ & $Q_L, L_L$\\
$ac'$ & $3\times (\overline{4},1,\overline{2},1,1)$ & -1 & 0 & $-1$   & $\frac 13,\; -\frac 23,\;1,\; 0$ & $-\frac 13,\;1$ & $Q_R, L_R$\\
$bc$ & $4\times (4,\overline{2},2,1,1)$ & 0 & -1 & $1$   & $-1,\; 0,\;0,\; 1$ & $0$ & $H$\\
$bc'$ & $8\times (4,\overline{2},\overline{2},1,1)$ & 0 & -1 & $-1$   & $-1,\; 0,\;0,\; 1$ & $0$ & $H'$\\
$bd$ & $2\times(1,2,1,\overline{2},1)$ & 0 & 1 & 0   & $\pm \frac 12$ & 0 & \\
$bd'$ & $1\times(1,\overline{2},1,\overline{2},1)$ & 0 & 1 & 0   & $\pm \frac 12$ & 0 & \\
$b4$ & $1\times(1,2,1,1,2)$ & 0 & 1 & 0   & $\mp \frac 12$ & 0 & \\
$cd$ & $3\times(1,1,\overline{2},2,1)$ & 0 & 0 & -1   & $\pm \frac 12$ & 0 & \\
$c4$ & $1\times(1,1,2,1,2)$ & 0 & 0 & 1   & $\pm \frac 12$ & 0 & \\
$b_{\yng(2)}$ & $2\times(1,3,1,1,1)$ & 0 & 2 & 0   & $0,\pm 1$ & 0 & \\
$b_{\overline{\yng(1,1)}}$ & $2\times(1,\overline{1},1,1,1)$ & 0 & -2 & 0   & 0 & 0 & \\
$c_{\yng(2)}$ & $3\times(1,1,3,1,1)$ & 0 & 0 & 3   & $0,\pm 1$ & 0 & \\
$c_{\overline{\yng(1,1)}}$ & $3\times(1,1,\overline{1},1,1)$ & 0 & 0 & -3   & 0 & 0 & \\
\hline
\end{tabular}
\end{center}
\end{table}

\subsection{Pati-Salam models with the gauge group ${\rm U}(4)_C \times {\rm USp}(2)_L \times {\rm SU}(2)_R$}\label{sec:Pheno1}
Three-family supersymmetric Pati-Salam with the gauge group ${\rm U}(4)_C \times {\rm USp}(2)_L \times {\rm SU}(2)_R$ have been recently discussed in ref.~\cite{Mansha:2022pnd}. It was found that the number of such models is only 5. However, by fixing the visible stacks $a$, $b$ and $c$, and searching for all possible hidden sectors results in two new models. The new models are Model~\hyperref[model2.1b]{2.1b} and Model~\hyperref[model6b]{5b}. In appendix \ref{appendix} we enlist all such three-family supersymmetric Pati-Salam models with a single symplectic group under the variation of the hidden sector. We refer readers to consult ref.~\cite{Mansha:2022pnd} for the detailed particle spectra.

\subsection{Pati-Salam models with the gauge group ${\rm U}(4)_C \times {\rm USp}(2)_L \times {\rm USp}(2)_R$}\label{sec:Pheno2}

\begin{widetext}  
	\begin{table}[htb]
		\centering \footnotesize
		\caption{D6-brane configurations and intersection numbers of Model~\hyperref[model2USp_1a]{1a}, and its gauge coupling relation is $g^2_a=F(x_B)g^2_b=F(x_B)g_c^2 = \frac{3 (3 x_B+8) x_B+8}{30 x_B+20}\,\left(\frac{5}{3}\,g^2_Y\right) 
			= \frac{4 \sqrt{2} (x_B(3 x_B+4))^{3/4}}{\sqrt[4]{3} (3 x_B+2)}\, \pi \,e^{\phi_4}$.\\} \label{model2USp_1a}
		$\begin{array}{|c||c|c| |r|r| r|r|r| r|r|r| r|r|r|}
			\hline
			\rm{Model~\hyperref[model2USp_1a]{1a}}  & \multicolumn{12}{c|}{{\rm U}(4)\times {\rm USp}(2)_L \times {\rm USp}(2)_R \times {\rm U}(2)\times {\rm USp}(8)}\\
			\hline \hline \rm{stack} & N & (n^1,l^1)\times (n^2,l^2)\times (n^3,l^3) & n_{\yng(2)} & n_{\yng(1,1)_{}} & b & c & d & d' & 1 & 2 & 3 & 4 \\
			\hline
			a     & 8     & (1, -3)\times (0, -1)\times (-3, 1) & 0     & 0       & 3           & -3          & 0           & 0     & 0     & 0     & 0     & 0 \\
			b     & 2     & (0, 1)\times (1, 0)\times (0, -2) & 0     & 0       & \text{-}    & 0           & 6          & -6    & 0     & 0     & 0     & 0 \\
			c     & 2     & (1, 0)\times (1, 0)\times (2, 0) & 0     & 0       & \text{-}    & \text{-}    & 6          & -6   & 0     & 0     & 0     & 0 \\
			d     & 4     & (1, -3)\times (1, -2)\times (3, 1) & -16    & -56      & \text{-}    & \text{-}    & \text{-}    & \text{-}    & 0    & 0  & 0   & 1 \\
			\hline
			4     & 8     & (0, -1)\times (0, 1)\times (2, 0) &  \multicolumn{10}{c|}{\beta^g_d=6, \quad\beta^g_4=-5,   } \\
			&       &                                   &   \multicolumn{10}{c|}{x_A=1=x_C, \quad x_D=9x_B+12 } \\
			&       &                                   &   \multicolumn{10}{c|}{\chi_1=x_B \chi_2=2/\chi_3 } \\
			\hline
		\end{array}$
	\end{table}
	\begin{table}[htb]
		\centering \footnotesize\renewcommand{\arraystretch}{1.3}
		\caption{The chiral and vector-like superfields, and their quantum numbers under the gauge symmetry ${\rm SU}(4)\times {\rm USp}(2)_L \times {\rm USp}(2)_R \times {\rm SU}(2) \times {\rm USp}(8)$ for the Model~\hyperref[model2USp_1a]{1a}.\\}\label{tab:2USp_1a}
		$\begin{array}{|c||c||r|r|r||c|c|c|}\hline
			\rm{Model~\hyperref[model2USp_1a]{1a}} & \text{Quantum Number}      & Q_4 & Q_{2L} & Q_{2R} & Q_{em} & B-L & \text{Field} \\
			\hline\hline
			ab               & 3 \times (4,\overline{2},1,1,1)           &  1  & -1  &  0  & -\frac{1}{3}, \frac{2}{3}, -1, 0 & \frac{1}{3}, -1  &  Q_L, L_L\\
			ac               & 3\times (\overline{4},1,2,1,1)            & -1  &  0  &  1  & \frac{1}{3}, -\frac{2}{3}, 1, 0  & -\frac{1}{3}, 1  &  Q_R, L_R\\
			bd               & 6\times (1,2,1,\overline{2},1)            &  0  &  1  &  0  &  \pm \frac{1}{2} & 0  &    \\
			bd'              & 6\times (1,\overline{2},1,\overline{2},1) &  0  & -1  &  0  &  \mp \frac{1}{2} & 0  &    \\
			cd               & 6\times (1,1,2,\overline{2},1)            &  0  &  0  &  1  &  \pm \frac{1}{2} & 0  &    \\
			cd'              & 6\times (1,1,\overline{2},\overline{2},1) &  0  &  0  & -1  &  \mp \frac{1}{2} & 0  &    \\
			d4               & 1\times (1,1,1,2,\overline{8})            &  0  &  0  &  0  &  0  &  0 &    \\
			d_{\overline{\yng(2)}}      & 16\times(1,1,1,\overline{3}_{\overline{\yng(2)}},1)            &  0 & 0 &  0  &  0  &  0 &    \\
			d_{\overline{\yng(1,1)}_{}} & 56\times(1,1,1,\overline{1}_{\overline{\yng(1,1)}_{}},1)            &  0 & 0 &  0  &  0  &  0 &    \\
			\hline\hline
			bc           & 2\times (1,\overline{2},2,1,1)                & 0 & -1 & 1  & 1, 0, 0, -1 &  0  &   H_u, H_d\\
			& 2\times (1,2,\overline{2},1,1)                & 0 &  1 & -1 &             &     &   \\
			\hline
		\end{array}$
	\end{table}
	
	\begin{table}[htb]
		\centering \footnotesize
		\caption{Model~\hyperref[model2USp_1b]{1b} represents the hidden sector variation of model III (T-dual) in ref.~\cite{Cvetic:2004nk}. D6-brane configurations and intersection numbers of Model~\hyperref[model2USp_1b]{1b}, and its gauge coupling relation is $g_a^2=\frac{13 g_b^2}{11}=\frac{13 g_c^2}{11}=\frac{61}{55} \frac{5 g_Y^2}{3}=\frac{8}{11} 13^{3/4}\, \pi \,e^{\phi_4}$.\\} \label{model2USp_1b}
		$\begin{array}{|c||c|c| |r|r| r|r|r| r|r|r| r|r|r|}
			\hline
			\rm{Model~\hyperref[model2USp_1b]{1b}}  & \multicolumn{10}{c|}{{\rm U}(4)\times {\rm USp}(2)_L \times {\rm USp}(2)_R \times {\rm U}(1)\times {\rm U}(1)}\\
			\hline \hline
			\rm{stack} & N & (n^1,l^1)\times (n^2,l^2)\times (n^3,l^3) & n_{\yng(2)} & n_{\yng(1,1)_{}} & b & c & d & d' & e & e' \\
			\hline
			a     & 8     & (1, -3)\times (0, -1)\times (-3, 1)  & 0     & 0      & 3        & -3       & 1        & 2        & -3       & 0  \\
			b     & 2     & (0, 1)\times (1, 0)\times (0, -2)    & 0     & 0      & 0        & 0        & 0        & 0        & 12       & -12  \\
			c     & 2     & (1, 0)\times (1, 0)\times (2, 0)     & 0     & 0      & \text{-} & \text{-} & 2        & -2       & 10       & -10  \\
			d     & 2     & (0, -1)\times (1, -2)\times (-1, 1)  & 1     & -1     & \text{-} & \text{-} & \text{-} & \text{-} & 0        & -8 \\
			e     & 2     & (2, -5)\times (-1, 2)\times (-3, -1) & -85   & -155   & \text{-} & \text{-} & \text{-} & \text{-} & \text{-} & \text{-} \\
			\hline
			&       &                                   &   \multicolumn{8}{c|}{\beta^g_d=-2, \quad \beta^g_e=22, }\\
			&       &                                   &   \multicolumn{8}{c|}{1/\chi_1=1/(2\chi_2)= \chi_3/2=\sqrt{13} } \\
			\hline
		\end{array}$
	\end{table}
	\begin{table}[htb] 
		\centering \footnotesize
		\caption{Model~\hyperref[model2USp_1c]{1c} represents the hidden sector variation of model III (T-dual) in ref.~\cite{Cvetic:2004nk}. D6-brane configurations and intersection numbers of Model~\hyperref[model2USp_1c]{1c}, and its gauge coupling relation is $g_a^2=\frac{92 g_b^2}{105}=\frac{92 g_c^2}{105}=\frac{162}{175} \frac{5 g_Y^2}{3}=\frac{32}{35} \left(\frac{23}{3}\right)^{3/4}\, \pi \,e^{\phi_4}$.\\} \label{model2USp_1c}
		$\begin{array}{|c||c|c| |r|r| r|r|r| r|r|r| r|r|r|}
			\hline
			\rm{Model~\hyperref[model2USp_1c]{1c}}  & \multicolumn{11}{c|}{{\rm U}(4)\times {\rm USp}(2)_L \times {\rm USp}(2)_R \times {\rm U}(1)\times {\rm U}(1) \times {\rm USp}(2)}\\
			\hline \hline
			\rm{stack} & N & (n^1,l^1)\times (n^2,l^2)\times (n^3,l^3) & n_{\yng(2)} & n_{\yng(1,1)_{}} & b & c & d & d' & e & e' & 4 \\
			\hline
			a     & 8     & (1, -3)\times (0, -1)\times (-3, 1)  & 0     & 0      & 3        & -3       & 0        & 3        & -3       & 0  & 0 \\
			b     & 2     & (0, 1)\times (1, 0)\times (0, -2)    & 0     & 0      & 0        & 0        & -3        & 3        & 9       & -9  & 0 \\
			c     & 2     & (1, 0)\times (1, 0)\times (2, 0)     & 0     & 0      & \text{-} & \text{-} & -4        & 4       & 6       & -6  &  0 \\
			d     & 2     & (-1, -4)\times (1, 1)\times (-3, 1)  & 6     & 42     & \text{-} & \text{-} & \text{-} & \text{-} & 72        & 0 &  -1\\
			e     & 2     & (-1, 2)\times (1, -3)\times (-3, -1) & -16   & -56   & \text{-} & \text{-} & \text{-} & \text{-} & \text{-} & \text{-} & 1 \\
			\hline
			4     & 2     & (0, -1)\times (0, 1)\times (2, 0) &   \multicolumn{9}{c|}{\beta^g_d=73, \quad \beta^g_e=87,\quad \beta^g_4=-4, }\\
			&       &                                   &   \multicolumn{9}{c|}{2/\chi_1=6\chi_2=\chi_3=\sqrt{69} } \\
			\hline
		\end{array}$
	\end{table}
	\begin{table}[htb]
		\centering \footnotesize\renewcommand{\arraystretch}{1.3}
		\caption{The chiral and vector-like superfields, and their quantum numbers under the gauge symmetry ${\rm SU}(4)\times {\rm USp}(2)_L \times {\rm USp}(2)_R \times {\rm U}(1)\times {\rm U}(1)$ for the Model~\hyperref[model2USp_1b]{1b}.\\}\label{tab:2USp_1b}
		$\begin{array}{|c||c||r|r|r||c|c|c|}\hline
			\rm{Model~\hyperref[model2USp_1b]{1b}} & \text{Quantum Number}      & Q_4 & Q_{2L} & Q_{2R} & Q_{em} & B-L & \text{Field} \\
			\hline\hline
			ab               & 3 \times (4,\overline{2},1,0,0)           &  1  & -1  &  0  & -\frac{1}{3}, \frac{2}{3}, -1, 0 & \frac{1}{3}, -1  &  Q_L, L_L\\
			ac               & 3\times (\overline{4},1,2,0,0)            & -1  &  0  &  1  & \frac{1}{3}, -\frac{2}{3}, 1, 0  & -\frac{1}{3}, 1  &  Q_R, L_R\\
			ad               & 1\times (4,1,1,-1,0)            &  1  &  0  &  0  & -\frac{1}{3}, \frac{2}{3}, -1, 0 & 0  &    \\
			ad'              & 2\times (4,1,1,1,0)                       &  1  &  0  &  0  & -\frac{1}{3}, \frac{2}{3}, -1, 0 & 0  &    \\
			ae               & 3\times (\overline{4},1,1,0,1)            & -1  &  0  &  0  & \frac{1}{3}, -\frac{2}{3}, 1, 0  & 0  &    \\
			be               & 12\times (1,2,1,0,1)           &  0  &  1  &  0  &  \pm \frac{1}{2} & 0  &    \\
			be'              & 12\times (1,\overline{2},1,0,-1)&  0  & -1  &  0  &  \mp \frac{1}{2} & 0  &    \\
			cd               & 2\times (1,1,2,-1,0)            &  0  &  0  &  1  &  \pm \frac{1}{2} & 0  &    \\
			cd'              & 2\times (1,1,\overline{2},-1,0) &  0  &  0  & -1  &  \mp \frac{1}{2} & 0  &    \\
			ce               & 10\times (1,1,2,0,-1)           &  0  &  0  &  1  &  \pm \frac{1}{2} & 0  &    \\
			ce'              & 10\times (1,1,\overline{2},0,-1)&  0  &  0  & -1  &  \mp \frac{1}{2} & 0  &    \\
			de'              & 8\times (1,1,1,-1,-1) &  0  &  0  &  0  &  0  &  0 &    \\
			d_{\yng(2)}      & 1\times(1,1,1,2_{\yng(2)},0)              &  0  &  0  &  0  &  0  &  0 &    \\
			e_{\overline{\yng(2)}}      & 85\times(1,1,1,0,-2_{\overline{\yng(2)}})             &  0  &  0  &  0  &  0  &  0 &    \\
			\hline\hline
			bc           & 2\times (1,\overline{2},2,1,1)                & 0 & -1 & 1  & 1, 0, 0, -1 &  0  &   H_u, H_d\\
			& 2\times (1,2,\overline{2},1,1)                & 0 &  1 & -1 &             &     &   \\
			\hline
		\end{array}$
	\end{table}

	\begin{table}[htb]
		\centering \footnotesize\renewcommand{\arraystretch}{1.3}
		\caption{The chiral and vector-like superfields, and their quantum numbers under the gauge symmetry ${\rm SU}(4)\times {\rm USp}(2)_L \times {\rm USp}(2)_R \times {\rm U}(1)\times {\rm U}(1) \times {\rm USp}(2)$ for the Model~\hyperref[model2USp_1c]{1c}.}\label{tab:2USp_1c}
		$\begin{array}{|c||c||r|r|r||c|c|c|}\hline
			\rm{Model~\hyperref[model2USp_1c]{1c}} & \text{Quantum Number}      & Q_4 & Q_{2L} & Q_{2R} & Q_{em} & B-L & \text{Field} \\
			\hline\hline
			ab               & 3\times (4,\overline{2},1,0,0,1)            &  1  & -1  &  0  & -\frac{1}{3}, \frac{2}{3}, -1, 0  & \frac{1}{3}, -1  &  Q_L, L_L\\
			ac               & 3\times (\overline{4},1,2,0,0,1)            & -1  &  0  &  1  & \frac{1}{3}, -\frac{2}{3}, 1, 0   & -\frac{1}{3}, 1  &  Q_R, L_R\\
			ad'              & 3\times (4,1,1,1,0,1)                       &  1  &  0  &  0  &  -\frac{1}{3}, \frac{2}{3}, -1, 0 & 0  &    \\
			ae               & 3\times (\overline{4},1,1,0,1,1)            & -1  &  0  &  0  &  \frac{1}{3}, -\frac{2}{3}, 1, 0  & 0  &    \\
			bd               & 3\times (1,\overline{2},1,1,0,1)            &  0  & -1  &  0  &  \mp \frac{1}{2} & 0  &    \\
			bd'              & 3\times (1,2,1,1,0,1)                       &  0  &  1  &  0  &  \pm \frac{1}{2} & 0  &    \\
			be               & 9\times (1,2,1,0,-1,1)            &  0  &  1  &  0  &  \pm \frac{1}{2} & 0  &    \\
			be'              & 9\times (1,\overline{2},1,0,-1,1) &  0  & -1  &  0  &  \mp \frac{1}{2} & 0  &    \\
			cd               & 4\times (1,1,\overline{2},1,0,1)            &  0  &  0  & -1  &  \mp \frac{1}{2} & 0  &    \\
			cd'              & 4\times (1,1,2,1,0,1)                       &  0  &  0  &  1  &  \pm \frac{1}{2} & 0  &    \\
			ce               & 6\times (1,1,2,0,-1,1)            &  0  &  0  &  1  &  \pm \frac{1}{2} & 0  &    \\
			ce'              & 6\times (1,1,\overline{2},0,-1,1) &  0  &  0  & -1  &  \mp \frac{1}{2} & 0  &    \\
			de               & 72\times (1,1,1,1,-1,1)           &  0  &  0  &  0  &  0  & 0  &    \\
			d4               & 1\times (1,1,1,-1,0,2)            &  0  &  0  &  0  &  0  &  0 &    \\
			e4               & 1\times (1,1,1,0,1,2)            &  0  &  0  &  0  &  0  &  0 &    \\
			d_{\yng(2)}      & 6\times(1,1,1,2_{\yng(2)},0,1)              &  0  &  0  &  0  &  0  &  0 &    \\
			e_{\overline{\yng(2)}}      & 16\times(1,1,1,0,-2_{\overline{\yng(2)}},1)             &  0  &  0  &  0  &  0  &  0 &    \\
			\hline\hline
			bc           & 2\times (1,\overline{2},2,1,1)                & 0 & -1 & 1  & 1, 0, 0, -1 &  0  &   H_u, H_d\\
			& 2\times (1,2,\overline{2},1,1)                & 0 &  1 & -1 &             &     &   \\
			\hline
		\end{array}$
	\end{table} 
\end{widetext}

Three-family supersymmetric Pati-Salam with the gauge group ${\rm U}(4)_C \times {\rm USp}(2)_L \times {\rm SU}(2)_R$ were first constructed in ref.~\cite{Cvetic:2004nk}. The authors of ref.~\cite{Cvetic:2004nk} (one of us, TL) only obtained a single consistent model with ${\cal N}=1$ supersymmetry with three generations. We reproduce the model in table~\ref{model2USp_1a} in its T-dual form \cite{Cvetic:2004ui} and its detailed particle spectrum is presented in table~\ref{tab:2USp_1a}. 
\begin{align}
	g^2_a =F(x_B)g^2_b=F(x_B)g_c^2 &= \frac{5 (3 x_B+4)}{2(3 x_B^2+13 x_B+6)}\,\left(\frac{5}{3}\,g^2_Y\right) 
	= \frac{4 \sqrt{2} (x_B(3 x_B+4))^{3/4}}{\sqrt[4]{3} (3 x_B+2)}\, \pi \,e^{\phi_4},\nonumber\\
	\intertext{where}
	F(x_B)&=\frac{x_B (3 x_B+4)}{2(3 x_B+2)} 
\end{align}
Choosing $F(x_B)=1$ by setting the value of free parameter $x_B=\frac{1}{3} \left(\sqrt{13}+1\right)$ the tree-level MSSM gauge couplings are unified at the string scale,
\begin{align}
	g^2_a = g^2_b= g^2_c=\left(\frac{5}{3}\,g^2_Y\right)= 4\ 2^{3/4} \sqrt[4]{\frac{1}{3} \left(\sqrt{13}-3\right)}\, \pi \,e^{\phi_4}.
\end{align}
Even though the gauge couplings are unified, this does not fix the actual value of the couplings as these still depend upon the value taken by the four-dimensional dilaton $\phi_4$. In order for the gauge couplings to have the value observed for the MSSM ($g^2_{\rm unification} \approx 0.511$), we must choose $\phi_4 = -3.32221$ such that $e^{-\phi_4} \approx 27.7216$, where the string scale is given by
\begin{equation}
	M_{St} = \pi^{1/2} e^{\phi_4} M_{Pl} \approx 1.55688 \times 10^{17}~\mbox{GeV},
\end{equation}
where $M_{Pl}$ is the reduced Planck scale. Obviously, there are one-loop threshold corrections arising from the ${\cal N}=1$ and ${\cal N}=2$ open string sectors~\cite{Lust:2003ky}. Additionally, there are exotic particles charged under both observable and hidden sector gauge groups, which are expected to pick up large masses, but could still affect the running of the gauge couplings.

Fixing the observable sectors $a$, $b$ and $c$ of Model~\hyperref[model2USp_1a]{1a} and varying the hidden sector we have found two additional models that are consistent with the supersymmetry conditions and the no-tadpole constraints. These additional new models are listed as Model~\hyperref[model2USp_1b]{1b} and Model~\hyperref[model2USp_1c]{1c} in tables~\ref{model2USp_1b} and \ref{model2USp_1c} respectively. The respective particle spectra of the models are also given in table~\ref{tab:2USp_1b} and table~\ref{tab:2USp_1c}.

In particular as can be noted from the descriptions of tables~\ref{model2USp_1b} and \ref{model2USp_1c} that unlike Model~\hyperref[model2USp_1a]{1a}, there is no free parameter in the gauge coupling relations of the models~\hyperref[model2USp_1b]{1b} and \hyperref[model2USp_1c]{1c}. However the gauge couplings still exhibit approximate gauge coupling unification at the string scale.


\section{String-scale Gauge coupling relations}\label{sec:RGE}
In this section, we will discuss the RGE running for the gauge couplings in the Model~\hyperref[model2USp_1b]{1b} and Model~\hyperref[model2USp_1c]{1c}.  The  RGEs for the gauge couplings at the two-loop level are given by \cite{Barger:2004sf,Barger:2007qb,Barger:2005qy,Gogoladze:2010in,Chen:2017rpn,Chen:2018ucf} 
{\small \begin{equation}
		\frac{d}{d\ln \mu} g_i=\frac{b_i}{(4\pi)^2}g_i^3 +\frac{g_i^3}{(4\pi)^4}
		\left[ \sum_{j=1}^3 B_{ij}g_j^2-\sum_{\alpha=u,d,e} d_i^\alpha
		{\rm Tr}\left( h^{\alpha \dagger}h^{\alpha}\right) \right],\label{eq:rge}
\end{equation}}
where $\mu$ is the running mass scale, $g_i(i=1,2,3)$ are the SM gauge couplings and $h^{\alpha}(\alpha=u,d,e)$ are the Yukawa couplings. To obtain the two-loop RGEs for SM gauge couplings, we perform numerically calculations including the one-loop RGEs for Yukawa couplings and taking into account the new physics contributions and threshold. And the whole integral is divided into three segments. The first one is from the electroweak scale, \textit{i.e.} Z boson mass scale $M_Z$, up to the supersymmetry breaking scale $M_S$, where we consider only the non-supersymmetric SM spectrum including a top quark pole mass at $m_t=173.34$ GeV and the corresponding gauge couplings at the scale $M_Z$ are
\begin{align}
	g_1(M_Z)=\sqrt{k_Y}\frac{g_{em}}{\cos\theta_W}\,,\quad
	g_2(M_Z)=\sqrt{k_2}\frac{g_{em}}{\sin\theta_W}\,,\quad 
	g_3(M_Z)=\sqrt{4\pi \alpha_s}\,.
\end{align}
The mass of Z boson is fixed at its experimental value $M_Z=91.1876 {\rm~GeV}$ in the following computations. The Higgs vacuum expectation value, strong coupling constant, fine structure constant, and weak mixing angle at $M_Z$ are choosen to be \cite{ParticleDataGroup:2018ovx,ParticleDataGroup:2020ssz}
\begin{gather}
	v=174.10 {\rm{~GeV}}\,,\quad 
	\sin^2\theta_W(M_Z)=0.23122\,,\quad \nonumber\\
	\alpha_s(M_Z)=0.1181\pm 0.0011\,,\quad 
	\alpha_{em}^{-1}(M_Z)=128.91\pm 0.02\,.
\end{gather} 
Next, we perform the supersymmetric RGEs from $M_{\rm S}$ scale, and the contributions form the introduced exotic vector-like particles are included from $M_V$ up to string scale. Thus, in the running of the gauge couplings, there can be two bending points corresponding to $M_S$ and $M_V$ and dividing the whole running lines into three region. In our calculation, the free inputs are the masses of these vector-like particles. Based on the experimental lower limits of supersymmetry and gauge hierarchy preservation, we have the supersymmetry breaking scale $M_{\rm S}$ at TeV scale. In our previous studies \cite{He:2021kbj,Mansha:2022pnd,Li:2022cqk}, we find that $M_{\rm S}$ within an order of magnitude gives a deviation on the scale of unification $M_{\rm U}$ less than $5\%$ and that the larger value for $M_{\rm S}$ reduces the unification scale. Thus in the following calculations, we set the supersymmetry breaking scale at $3.0$ TeV. At last, we get the gauge coupling relations at the string scale as 
\begin{equation}
	g_a^2=k_2g_b^2=k_Y g_Y^2 =g_{\text{U}}^2 \sim g_{\text{string}}^2~,~\,
\end{equation}
where $g_a$, $g_b$, and $g_Y$ are respectively the gauge couplings for ${\rm SU}(3)_C$, ${\rm SU}(2)_L$, and ${\rm U}(1)_Y$. $k_2$ and $k_Y$ are rational numbers. The canonical normalization in ${\rm SU}(5)$ and ${\rm SO}(10)$ models give $k_2=1$ and $k_Y=5/3$.

The coefficients of beta functions in Eq. \eqref{eq:rge} in SM \cite{Machacek:1983tz,Machacek:1983fi,Machacek:1984zw,Cvetic:1998uw} and supersymmetric models \cite{Barger:1992ac,Barger:1993gh,Martin:1993zk} are represented by
\begin{align}
	b_{\rm SM}&=\left(\frac{41}{6} \frac{1}{k_Y},-\frac{19}{6}\frac{1}{k_2},-7\right) ,\quad 
	B_{\rm SM}=\begin{pmatrix}
		\frac{199}{18} \frac{1}{k_Y^2} &
		\frac{27}{6} \frac{1}{k_Y k_2} &\frac{44}{3} \frac{1}{k_Y} \cr 
		\frac{3}{2} \frac{1}{k_Y k_2} & \frac{35}{6}\frac{1}{k_2^2}&12\frac{1}{k_2} \cr
		\frac{11}{6} \frac{1}{k_Y} &\frac{9}{2}\frac{1}{ k_2}&-26 
	\end{pmatrix},\quad \\
	d^u_{\rm SM}&=\left(\frac{17}{6} \frac{1}{k_Y} ,\frac{3}{2}\frac{1}{k_2},2\right),\quad
	d^d_{\rm SM}=0,\quad
	d^e_{\rm SM}=0, \,\\
	b_{\rm SUSY}&=\left(11 \frac{1}{k_Y},\frac{1}{k_2},-3\right) ,\quad 
	B_{\rm SUSY}=
	\begin{pmatrix}
		\frac{199}{9}
		\frac{1}{k_Y^2}&  9\frac{1}{k_Yk_2}&\frac{88}{3} \frac{1}{k_Y} \cr
		3\frac{1}{k_Yk_2} & 25\frac{1}{k_2^2}&24\frac{1}{k_2} \cr
		\frac{11}{3}\frac{1}{k_Y} & 9\frac{1}{k_2} & 14
	\end{pmatrix},~ \\
	d^u_{\rm SUSY}&=\left(\frac{26}{3} \frac{1}{k_Y},6\frac{1}{k_2},4\right) ,\quad
	d^d_{\rm SUSY}=0,\quad 
	d^e_{\rm SUSY}=0,
\end{align}
where $k_Y$ and $k_2$ are general normalization factors. The general one-loop RGEs for Yukawa couplings can be found in~\cite{Gogoladze:2010in}.

\begin{figure}
	\centering
	\includegraphics[width=0.45\linewidth]{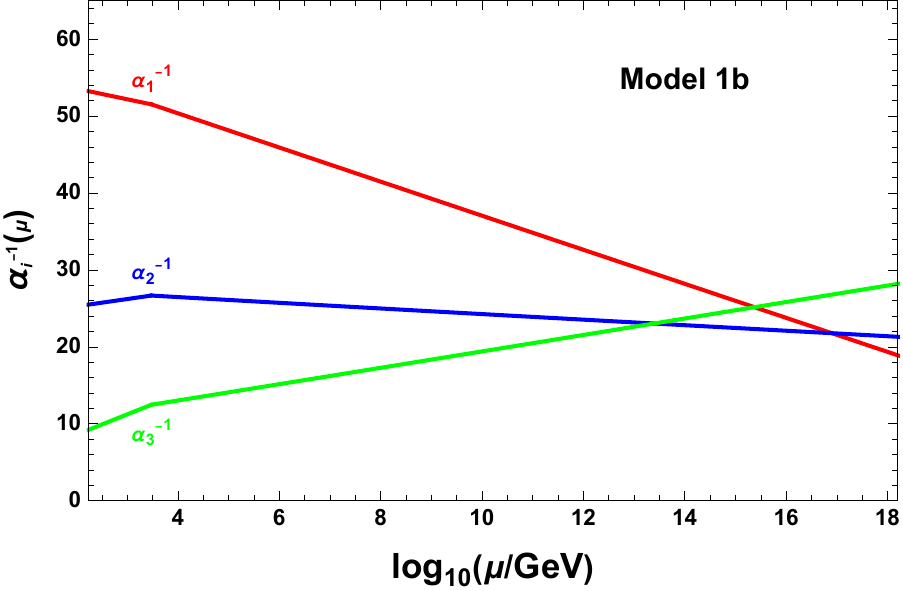}\quad
	\includegraphics[width=0.45\linewidth]{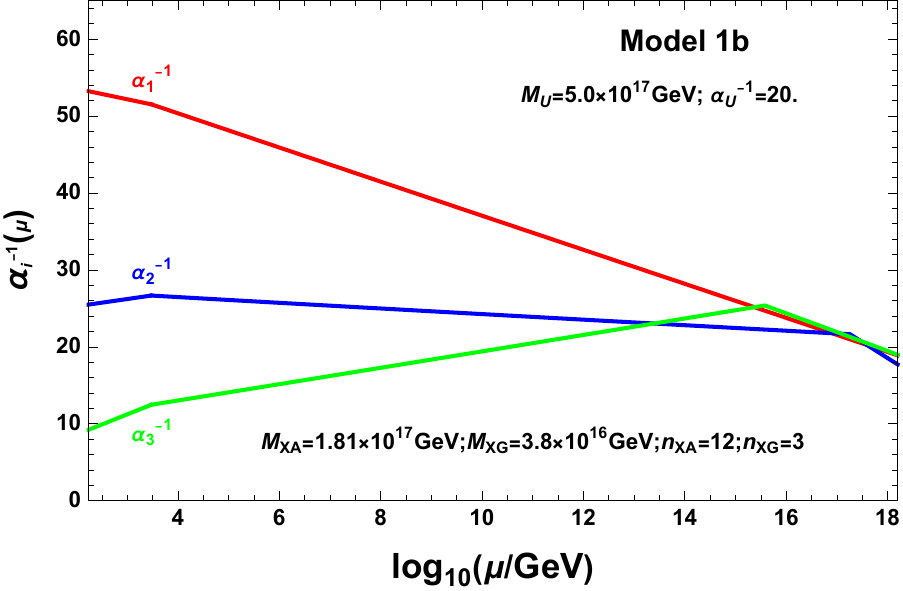}
	\caption{Two-loop evolution of gauge couplings for the Model ~\hyperref[model2USp_1b]{1b} without (left) and with (right) $12(XA+\overline{XA})$ at $1.81\times 10^{17}$ GeV and $3XG$ at $3.8\times10^{16}$ GeV. }\label{fig:rge-1b}
\end{figure}

\begin{figure}
	\includegraphics[width=0.45\linewidth]{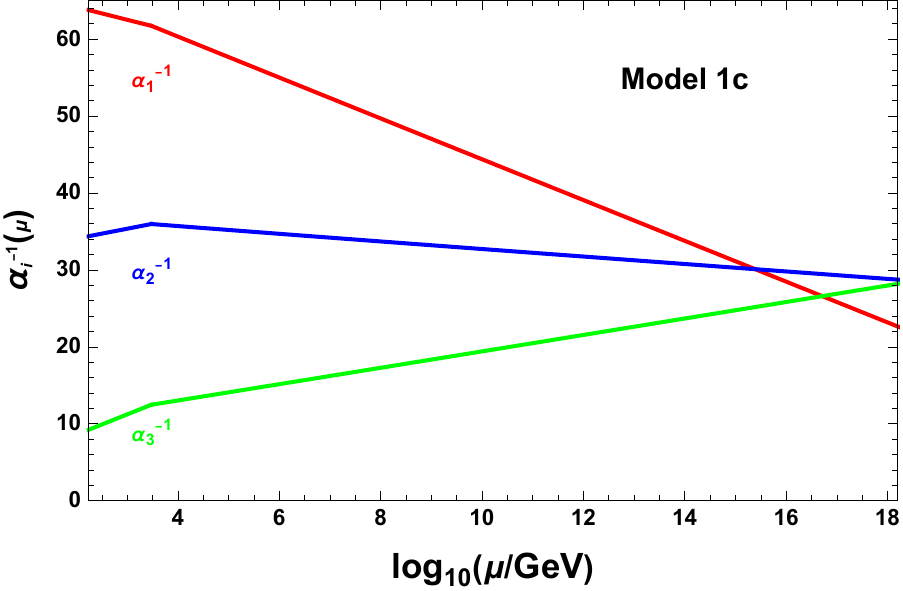}\quad
	\includegraphics[width=0.45\linewidth]{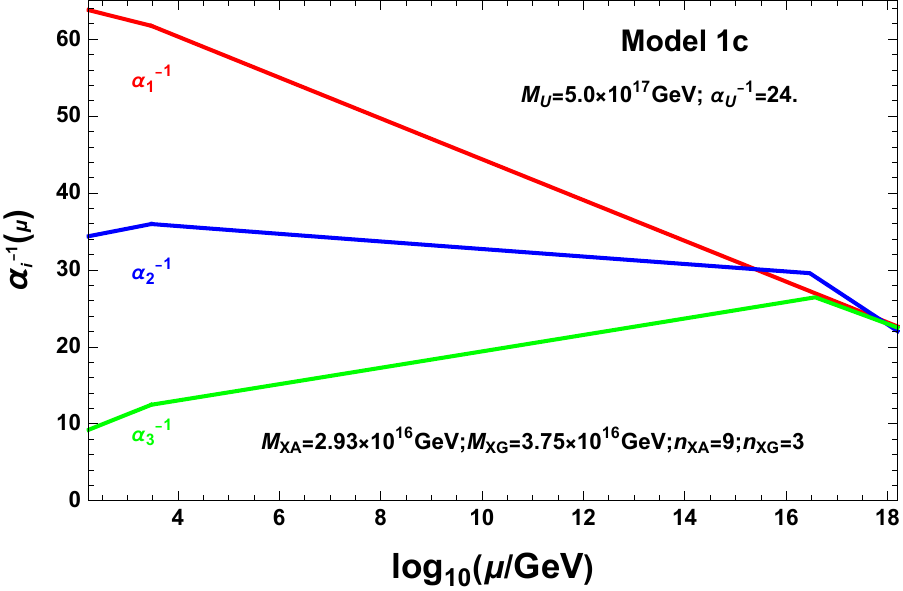}
	\caption{Two-loop evolution of gauge couplings for the Model ~\hyperref[model2USp_1c]{1c} without (left) and with (right) $9(XA+\overline{XA})$ at $2.93\times 10^{16}$ GeV and $3XG$ at $3.75\times10^{16}$ GeV. }
	\label{fig:rge-1c}
\end{figure}

Defining the parameters as
\begin{equation}
	\alpha_1\equiv k_Yg_Y^2/4\pi, \quad \alpha_2 \equiv k_2g_b^2/4\pi, \quad \alpha_3 \equiv g_a^2/4\pi,
\end{equation}
the two-loop evolution of gauge couplings  are shown in Figs.\ref{fig:rge-1b} and \ref{fig:rge-1c} for the models \hyperref[model2USp_1b]{1b} and \hyperref[model2USp_1c]{1c}, where the string-scale gauge coupling relation is achieved by setting $\alpha_{\text{U}}^{-1}\equiv \alpha_1^{-1}=(\alpha_2^{-1}+\alpha_3^{-1})/2$ and limiting the accuracy  $\Delta=|\alpha_1^{-1}-\alpha_2^{-1}|/\alpha_1^{-1}\leq 0.1\%$. The string-scale gauge coupling relations are realized by adding the vector-like particles $XA+ {\overline{XA}}$ from the fundamental representation of $SU(4)$ gauge group and the particle $XG$ from the adjoint representation of $SU(2)$. The quantum numbers of these exotic particles under $SU(3)_C\times SU(2)_L\times U(1)_Y$ are $XA + {\overline{XA}} = {\mathbf{(1,2, {0}) + (1,{\bar 2}, 0)}}$ and $XG=\mathbf{(8,1,0)}$. For the particles $XA + {\overline{XA}}$, the non-zero coefficients of one- and two-loop beta functions in the supersymmetric models are $\Delta b_2= 1$ and $\Delta B_{22}=7$, which will only modify the evolution of the electroweak coupling; while the the non-zero contributions from particle $XG$ are $\Delta b_3=3$ and $\Delta B_{33}=54$, which will alter the evolution of the strong coupling. Thus, the adding of these particles will bend the values of $\alpha_2^{-1}$ and $\alpha_3^{-1}$ to achieve the unification on the right panel of Figs. \ref{fig:rge-1b} and \ref{fig:rge-1c}.  Based on the intersection of D-branes, $12$ and $9$ pairs of $XA+\overline{XA}$, from the $be+be'$ sectors in the spectrum Tables~\ref{tab:2USp_1b} and \ref{tab:2USp_1c}, are added in the models \hyperref[model2USp_1b]{1b} and \hyperref[model2USp_1c]{1c}, respectively. While, the particle $XG$ arises from the $aa$ sectors, and in our calculations we introduce the particle $XG$ with the maximum number 3.

\section{Machine Learning}\label{sec:ml}
\begin{figure}[!h]
    \centering
    \includegraphics{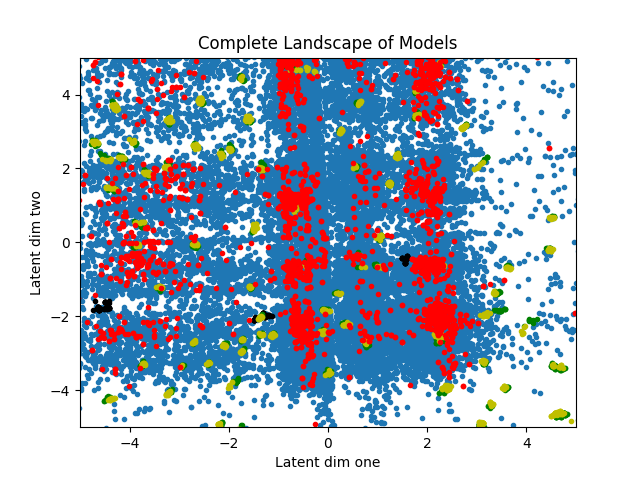}
    \caption{Landscape of Pati-Salam models where each point corresponds to a model: blue, and red points represent non-MSSM models and non-MSSM models with at least one USp group respectively, while green and yellow points represent MSSM models, and their hidden sector variation. Black points represent models with at least one USp group under the hidden sector variation. Non-MSSM models (with or without USp group) tend to overlap in clusters. MSSM models (with their hidden sector variation) overlap, and occupy separate regions. So is the models with USp group under the hidden sector variation.}
    \label{fig:landscape}
\end{figure}

We perform machine learning on scanned data with the help of autoencoder to visualise the models in form of a point in two-dimensions as depicted in Fig. \ref{fig:landscape}. We choose mean squared error (MSE) as the loss function. The gradients of loss to the learnable parameters are evaluated on each batch of size 500 examples and the learnable
parameters are optimized by using the ADAM optimizer \cite{kingma2017adam} with learning rate of 0.01.

Since the scan over the hidden sector is performed by fixing the visible sector, the complex structure moduli would remain unchanged. Thus, all the gauge coupling relations would remain same under the variation of the hidden sector except that the USp groups from the hidden sector change to unitary gauge groups. This leads to the generations of phenomenologically different models, but with same gauge coupling relations. It can also be verified from Fig. \ref{fig:landscape} where Pati-Salam models and their corresponding hidden sector variation (shown by green and yellow points respectively) approximately overlap, and sit in separate regions.

\begin{figure}[!h]
    \centering
    \includegraphics{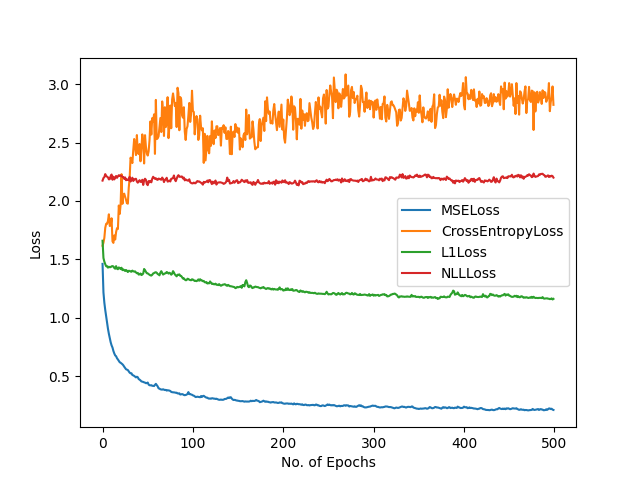}
    \caption{A comparison of different loss functions during training of data}
    \label{fig:loss}
\end{figure}

We use different loss functions to train the data on autoencoder, and show their efficiency in Fig.~\ref{fig:loss}.
It is clear from the Fig.~\ref{fig:loss} that loss is minimized through back-propagation in case of MSE loss function.

\section{Conclusion}\label{sec:conclusion}
We have found new three-family supersymmetric Pati-Salam models where either one or both of the ${\rm SU}(2)_{L,R}$ gauge factors arise from an original ${\rm USp}(2)_{L,R}$ group on a $\mathbbm{T}^6/(\mathbbm{Z}_2\times \mathbbm{Z}_2)$ orientifold from intersecting D6-branes at angles from IIA string theory without fluxes. Unlike the special unitary group, the symplectic group is a simpler choice as there is no associated global anomalous U(1) group. The unitary symplectic group arises by placing the D6-brane stack parallel to any of the four O6-planes.

By fixing the observable sector while varying the hidden sector we are able to ascertain the all possible gauge group factors consistent with the no-tadpole constraint. We have displayed the perturbative particle spectra of these representative models where the exotic particles are found to be decoupled in some of the models. This seemingly minor change from SU(2) to USp(2) group is quite restrictive and drastically reduces the number of consistent three-family models. Therefore, while the number of three-family models with ${\rm SU}(4)_C \times {\rm SU}(2)_L \times {\rm SU}(2)_R$ groups is found to be 202,752~\cite{He:2021kbj, He:2021gug}, the similar models by replacing either one of the SU(2) factors with a USp(2) is only 5~\cite{Mansha:2022pnd}. Interestingly, three-family supersymmetric Pati-Salam model with the specific gauge group ${\rm SU}(4)_C \times {\rm USp}(2)_L \times {\rm USp}(2)_R$ are even more constrained and we have only found two such models.

Finally, we also point out a discrepancy in ref.~\cite{He:2021gug} where a complete search of the landscape of supersymmetric Pati-Salam models constructed from Type IIA string theory on $\mathbbm{T}^6/(\mathbbm{Z}_2\times \mathbbm{Z}_2)$ orientifolds was claimed. Here, we have demonstrated that in order to chart out the complete landscape the inequivalent models with similar visible sectors but having a different hidden sector should also be taken into account as it can not only change the number of consistent models but also can affect the gauge coupling relations.

\begin{acknowledgments}
	We would like to thank Mudassar Sabir very much for collaboration in the initial stage of the project and helpful discussions. This research is supported in part by the National Key Research and Development Program of China Grant No. 2020YFC2201504, by the Projects No. 11875062, No. 11947302, No. 12047503, and No. 12275333 supported by the National Natural Science Foundation of China, by the Key Research Program of the Chinese Academy of Sciences, Grant No. XDPB15, by the Scientific Instrument Developing Project of the Chinese Academy of Sciences, Grant No. YJKYYQ20190049, and by the International Partnership Program of Chinese Academy of Sciences for Grand Challenges, Grant No. 112311KYSB20210012.
\end{acknowledgments}

\FloatBarrier

\bibliographystyle{apsrev4-2}
\bibliography{PS-USp_x_USp.bib}

\appendix

\section{Pati-Salam models with a symplectic group under the variations of the hidden sector}\label{appendix}

In this appendix, we list all representative three-family supersymmetric Pati-Salam models with a symplectic group under the variations of the hidden sector obtained from random scanning method. $a, b, c, d$ in the first column in every table represent the four stacks of D6-branes, respectively. Similarly, $1, 2, 3, 4$ in the first columns is a short-handed notation for the filler branes along the $\Omega R$, $\Omega R \omega$, $\Omega R \theta \omega$ and $\Omega R \theta$ O6-planes, respectively. The second column in each table lists the numbers of D6-branes in the respective stack. In the third column we record the wrapping numbers of each D6-brane configuration. The rest of the columns record the intersection numbers between various stacks. For instance, in the $b$ column of table~\ref{model0}, from top to bottom, the numbers represent intersection numbers $I_{ab}, I_{bc}, I_{bd}$, etc.  As usual, $b'$ and $c'$ are the orientifold $\Omega R$ image of $b$ and $c$ stacks of D6-branes. We also list the relation between $x_A, x_B, x_C, x_D$, which are determined by the supersymmetry conditions~\eqref{susyconditions}, as well as the relation between the moduli parameter $\chi_1,\, \chi_2,\, \chi_3$. The one loop beta functions $\beta^g_i$ for each of the hidden sector stack is also listed. The gauge coupling relations are given in the caption of each table.

\begin{widetext} \begin{table}[htb]  
		\centering \footnotesize
		\caption{Model~\hyperref[model0]{1} represents the hidden sector variation of Model 1 in ref.~\cite{Mansha:2022pnd}. D6-brane configurations and intersection numbers of Model~\hyperref[model0]{1}, and its gauge coupling relation is
			$g^2_a=\frac{8}{7}\, g^2_b=\frac{4}{3}\, g_c^2= \frac{6}{5}\,\left(\frac{5}{3}\,g^2_Y\right)= \frac{32}{7}\sqrt[4]{\frac{2}{3}}\, \pi \,e^{\phi_4}$.\\} \label{model0}
		$\begin{array}{|c|c|c |r|r|r| r|r|r| r|r|r| r|r|r|} 	\hline 	\multicolumn{3}{|c|}{\text{Model~\hyperref[model0]{1}}}  & \multicolumn{11}{c|}{{\rm U}(4)\times {\rm USp}(2)_L \times {\rm U}(2)_R \times {\rm U}(2)} \\
			\hline \hline \rm{stack} & N & (n^1,l^1)\times (n^2,l^2)\times (n^3,l^3) & n_{\yng(2)} & n_{\yng(1,1)_{}} & b & c & c'& d & d' & 1 & 2 & 3 & 4 \\
			\hline
			a   & 8   & (-1, -3)\times (0, -1)\times (-1, -1) & -2    & 2     & 3    & -3    & 0     & 4     & 4     & 0     & 0     & 0    & 0 \\
			b   & 2   & (1, 0)\times (1, 0)\times (2, 0)      & 0     & 0     & \text{-} & 4     & -4    & 2     & -2    & 0     & 0     & 0    & 0 \\
			c   & 4   & (0, 1)\times (3, -4)\times (1, -1)    & 1     & -1    & \text{-} & \text{-}  & \text{-}  & 4     & -10   & 0     & 0     & 0    & 0 \\
			d   & 4   & (-1, 1)\times (1, -2)\times (-3, -1)  & -2    & -22   & \text{-} & \text{-}  & \text{-}  & \text{-}  & \text{-}  & 0     & 0     & 0    & 0 \\
			\hline
			&     &                     &  \multicolumn{11}{c|}{\beta^g_{d}=8, \quad  x_A=\frac{3}{4}x_B=\frac{1}{3}x_C=\frac{1}{24}x_D} \\
			&     &                     &  \multicolumn{11}{c|}{18\chi_1=8\chi_2=\chi_3/2=\sqrt{6}} \\
			\hline
		\end{array}$
	\end{table}
	
	\begin{table}[htb]  
		\centering \footnotesize
		\caption{Model~\hyperref[model2.1a]{2a} represents the hidden sector variation of Model 2 in ref.~\cite{Mansha:2022pnd}. D6-brane configurations and intersection numbers of Model~\hyperref[model2.1a]{2a}, and its gauge coupling relation is $g^2_a=\frac{11}{36}\, g^2_b=\frac{1}{3}\, g_c^2= \frac{3}{5}\,\left(\frac{5}{3}\,g^2_Y\right)= \frac{2 \sqrt{2}}{9} 11^{3/4}\, \pi \,e^{\phi_4}$.\\} \label{model2.1a}
		$\begin{array}{|c|c|c |r|r|r| r|r|r| r|r|r| r|r|r|}
			\hline 	\multicolumn{3}{|c|}{\rm{Model~\hyperref[model2.1a]{2a}}}  & \multicolumn{11}{c|}{{\rm U}(4)\times {\rm USp}(2)_L \times {\rm U}(2)_R \times {\rm U}(2) \times {\rm USp}(2)}\\
			\hline \hline \rm{stack} & N & (n^1,l^1)\times (n^2,l^2)\times (n^3,l^3) & n_{\yng(2)} & n_{\yng(1,1)_{}} & b & c & c'& d & d' & 1 & 2 & 3 & 4 \\
			\hline
			a     & 8   & (-1, -3)\times (-1, 0)\times (1, -1) & -2    & 2     & 3        & 0        & -3       & 4     & 4           & 0     & 0     & 0     & 0 \\
			b     & 2   & (1, 0)\times (0, 1)\times (0, -2)    & 0     & 0     & \text{-} & -1       & 1        & 1     & -1          & 0     & 0     & 0     & 0 \\
			c     & 4   & (0, 1)\times (1, -3)\times (1, -1)   & 2     & -2    & \text{-} & \text{-} & \text{-} & 4     & -4          & -3    & 0     & 0     & 0 \\
			d     & 4   & (-1, 1)\times (-1, -1)\times (1, -3) & 0     & -12   & \text{-} & \text{-} & \text{-} & \text{-} & \text{-} & -3    & 0     & 0     & 0 \\
			\hline
			1       & 2   & (1, 0)\times (1, 0)\times (2, 0) &  \multicolumn{11}{c|}{\beta^g_{d}=7,~ \beta^g_1=0, } \\
			&     &                                  &  \multicolumn{11}{c|}{  x_A=\frac{1}{3}x_B=\frac{1}{33}x_C=\frac{1}{9}x_D }  \\
			&     &                                  &  \multicolumn{11}{c|}{33\chi_1=3\chi_2=\frac{11}{2} \chi_3= \sqrt{11} }  \\
			\hline
		\end{array}$
	\end{table}

	\begin{table}[htb]  
		\centering \footnotesize
		\caption{Model~\hyperref[model2.1b]{2b} represents the hidden sector variation of Model 2 in ref.~\cite{Mansha:2022pnd}. D6-brane configurations and intersection numbers of Model~\hyperref[model2.1b]{2b}, and its gauge coupling relation is $g^2_a=\frac{11}{36}\, g^2_b=\frac{1}{3}\, g_c^2= \frac{3}{5}\,\left(\frac{5}{3}\,g^2_Y\right)= \frac{2 \sqrt{2}}{9} 11^{3/4}\, \pi \,e^{\phi_4}$.\\} \label{model2.1b}
		$\begin{array}{|c|c|c |r|r|r| r|r|r| r|r|r| r|r|r|}
			\hline 	\multicolumn{3}{|c|}{\rm{Model~\hyperref[model2.1b]{2b}}}  & \multicolumn{11}{c|}{{\rm U}(4)\times {\rm USp}(2)_L \times {\rm U}(2)_R \times {\rm U}(2) \times {\rm USp}(6)}\\
			\hline \hline \rm{stack} & N & (n^1,l^1)\times (n^2,l^2)\times (n^3,l^3) & n_{\yng(2)} & n_{\yng(1,1)_{}} & b & c & c'& d & d' & 1 & 2 & 3 & 4 \\
			\hline
			a     & 8   & (-1, -3)\times (-1, 0)\times (1, -1) & -2    & 2     & 3        & 0        & -3       & 4     & 4           & 0     & 0     & 0     & 0 \\
			b     & 2   & (1, 0)\times (0, 1)\times (0, -2)    & 0     & 0     & \text{-} & -1       & 1        & -2     & 2          & 0     & 0     & 0     & 0 \\
			c     & 4   & (0, 1)\times (1, -3)\times (1, -1)   & 2     & -2    & \text{-} & \text{-} & \text{-} & -10     & 7          & 0    & 0     & 0     & 0 \\
			d     & 4   & (1, 1)\times (2, -1)\times (-1, 3)   & 2     & 22   & \text{-} & \text{-} & \text{-} & \text{-} & \text{-} & 0    & 0     & 0     & -1 \\
			\hline
			3       & 6   & (0, -1)\times (1, 0)\times (0, 2) &  \multicolumn{11}{c|}{\beta^g_{d}=14,~ \beta^g_3=-5, } \\
			&     &                                  &  \multicolumn{11}{c|}{x_A=\frac{1}{3}x_B=\frac{1}{63}x_C=\frac{1}{9}x_D  }  \\
			&     &                                  &  \multicolumn{11}{c|}{63\chi_1=3\chi_2=\frac{21}{2} \chi_3= \sqrt{21}   }  \\
			\hline
		\end{array}$
	\end{table}

	\begin{table}[htb]  
		\centering \footnotesize
		\caption{Model~\hyperref[model2.2]{3} represents the hidden sector variation of Model 3 in ref.~\cite{Mansha:2022pnd}. D6-brane configurations and intersection numbers of Model~\hyperref[model2.2]{3}, and its gauge coupling relation is $g^2_a=\frac{5}{18}\, g^2_b=\frac{1}{3}\, g_c^2= \frac{3}{5}\,\left(\frac{5}{3}\,g^2_Y\right)= \frac{4 \sqrt{2}}{9} 5^{3/4} \, \pi \,e^{\phi_4}$.\\} \label{model2.2}
		$\begin{array}{|c|c|c |r|r|r| r|r|r| r|r|r| r|r|r|}
			\hline 	\multicolumn{3}{|c|}{\rm{Model~\hyperref[model2.2]{3}}}  & \multicolumn{11}{c|}{{\rm U}(4)\times {\rm USp}(2)_L \times {\rm U}(2)_R \times {\rm U}(2) \times {\rm USp}(2)}\\
			\hline \hline \rm{stack} & N & (n^1,l^1)\times (n^2,l^2)\times (n^3,l^3) & n_{\yng(2)} & n_{\yng(1,1)_{}} & b & c & c'& d & d' & 1 & 2 & 3 & 4 \\
			\hline
			a     & 8     & (-1, 0)\times (-1, -3)\times (1, -3) & 0     & 0       & 3     & -3    & 0     & 0     & 0     & 0     & 0     & 0     & 0 \\
			b     & 2     & (0, 1)\times (1, 0)\times (0, -2) & 0     & 0         & \text{-}    & 1     & -1    & 6     & -6    & 0     & 0     & 0     & 0 \\
			c     & 4     & (1, 1)\times (0, 1)\times (-1, -3) & -2    & 2         & \text{-}    & \text{-}    & \text{-}    & 3     & 0     & 0     & 0     & 0     & 0 \\
			d     & 4     & (2, 1)\times (1, -3)\times (1, -3) & -16   & -56      & \text{-}    & \text{-}    & \text{-}    & \text{-}    & \text{-}    & 0     & 1     & 0     & 0 \\
			\hline
			2     & 2     & (1, 0)\times (0, -1)\times (0, 2) &  \multicolumn{11}{c|}{\beta^g_{d}=3,~ \beta^g_2=-5, } \\
			&       &                                   &  \multicolumn{11}{c|}{ x_A=\frac{1}{45}x_B=\frac{1}{3}x_C=\frac{1}{3}x_D}  \\
			&       &                                   &  \multicolumn{11}{c|}{\chi_1=15\chi_2=\frac{15}{2}\chi_3= \sqrt{5}}  \\
			\hline
		\end{array}$
	\end{table}
	
	\begin{table}[htb] 
		\centering \footnotesize
		\caption{Model~\hyperref[model4]{4} represents the hidden sector variation of Model 4 in ref.~\cite{Mansha:2022pnd}. D6-brane configurations and intersection numbers of Model~\hyperref[model4]{4}, and its gauge coupling relation is $g^2_a=\frac{11}{18}\, g^2_b=\frac{2}{3}\, g_c^2= \frac{4}{5}\,\left(\frac{5}{3}\,g^2_Y\right)= \frac{4}{9}\, 11^{3/4}\, \pi \,e^{\phi_4}$.\\} \label{model4}
		$\begin{array}{|c|c|c |r|r|r| r|r|r| r|r|r| r|r|r|}
			\hline 	\multicolumn{3}{|c|}{\rm{Model~\hyperref[model4]{4}}}  & \multicolumn{11}{c|}{{\rm U}(4)\times {\rm USp}(2)_L \times {\rm U}(2)_R \times {\rm U}(2) \times {\rm USp}(4)}\\
			\hline \hline \rm{stack} & N & (n^1,l^1)\times (n^2,l^2)\times (n^3,l^3) & n_{\yng(2)} & n_{\yng(1,1)_{}} & b  & c & c'& d & d' & 1 & 2 & 3 & 4 \\
			\hline
			a     & 8     & (-1, 0)\times (-1, -3)\times (1, -1) & -2    & 2      & 3     & -3    & 0     & 4     & 4     & 0     & 0     & 0     & 0 \\
			b     & 2     & (0, 1)\times (1, 0)\times (0, -2) & 0     & 0         & \text{-}    & 2     & -2    & 2     & -2    & 0     & 0     & 0     & 0 \\
			c     & 4     & (2, 3)\times (0, 1)\times (-1, -1) & -1    & 1         & \text{-}    & \text{-}    & \text{-}    & 8     & -8    & 0     & 0     & 0     & 0 \\
			d     & 4     & (2, 1)\times (1, -1)\times (1, -3) & -2    & -22      & \text{-}    & \text{-}    & \text{-}    & \text{-}    & \text{-}    & 0     & 1     & 0     & 0 \\
			\hline
			2     & 4     & (1, 0)\times (0, -1)\times (0, 2) &  \multicolumn{11}{c|}{\beta^g_{d}=12,~ \beta^g_2=-5,  } \\
			&       &                      &   \multicolumn{11}{c|}{ x_A=\frac{1}{33}x_B=\frac{2}{3}x_C=\frac{2}{9}x_D}  \\
			&       &                      &   \multicolumn{11}{c|}{3\chi_1=33\chi_2=\frac{11}{2}\chi_3= \sqrt{11} }  \\
			\hline
		\end{array}$
	\end{table}
	
	\begin{table}[htb] 
		\centering \footnotesize
		\caption{Model~\hyperref[model6a]{5a} represents the hidden sector variation of Model 5 in ref.~\cite{Mansha:2022pnd}. D6-brane configurations and intersection numbers of Model~\hyperref[model6a]{5a}, and its gauge coupling relation is $g^2_a=\frac{7}{22}\, g^2_b=\frac{1}{3}\, g_c^2= \frac{3}{5}\,\left(\frac{5}{3}\,g^2_Y\right)= \frac{4 \sqrt{2}}{11}\frac{7^{3/4}}{\sqrt[4]{3}}\, \pi \,e^{\phi_4}$.\\} \label{model6a}
		$\begin{array}{|c|c|c |r|r|r| r|r|r| r|r|r| r|r|r|}
			\hline 	\multicolumn{3}{|c|}{\rm{Model~\hyperref[model6a]{5a}}}  & \multicolumn{11}{c|}{{\rm U}(4)\times {\rm USp}(2)_L \times {\rm U}(2)_R \times {\rm U}(2)\times {\rm USp}(6)}\\
			\hline \hline \rm{stack} & N & (n^1,l^1)\times (n^2,l^2)\times (n^3,l^3) & n_{\yng(2)} & n_{\yng(1,1)_{}} & b & c & c'& d & d' & 1 & 2 & 3 & 4 \\
			\hline
			a     & 8     & (1, 3)\times (1, 0)\times (1, -1) & -2    & 2       & 3     & 0     & -3    & 4     & 4     & 0     & 0     & 0     & 0 \\
			b     & 2     & (1, 0)\times (0, 1)\times (0, -2) & 0     & 0       & \text{-}    & -1    & 1     & 2     & -2    & 0     & 0     & 0     & 0 \\
			c     & 4     & (0, -1)\times (-1, 3)\times (1, -1) & 2     & -2       & \text{-}    & \text{-}    & \text{-}    & 7     & -10   & 0     & 0     & 0     & 0 \\
			d     & 4     & (1, -1)\times (2, 1)\times (1, -3) & -2    & -22      & \text{-}    & \text{-}    & \text{-}    & \text{-}    & \text{-}    & 0     & 0     & 1     & 0 \\
			\hline
			3     & 6     & (0, -1)\times (1, 0)\times (0, 2) &  \multicolumn{11}{c|}{\beta^g_{d}=11,~ \beta^g_3=-5, } \\
			&       &                                   &   \multicolumn{11}{c|}{  x_A=\frac{1}{3}x_B=\frac{1}{63}x_C=\frac{1}{9}x_D}  \\
			&       &                                   &   \multicolumn{11}{c|}{63\chi_1=3\chi_2=\frac{21}{2}\chi_3= \sqrt{2} }  \\
			\hline
		\end{array}$
	\end{table}
	
	\begin{table}[htb] 
		\centering \footnotesize
		\caption{Model~\hyperref[model6b]{5b} represents the hidden sector variation of Model 5 in ref.~\cite{Mansha:2022pnd}. D6-brane configurations and intersection numbers of Model~\hyperref[model6b]{5b}, and its gauge coupling relation is $g^2_a=\frac{7}{22}\, g^2_b=\frac{1}{3}\, g_c^2= \frac{3}{5}\,\left(\frac{5}{3}\,g^2_Y\right)= \frac{4 \sqrt{2}}{11}\frac{7^{3/4}}{\sqrt[4]{3}}\, \pi \,e^{\phi_4}$.\\} \label{model6b}
		$\begin{array}{|c|c|c |r|r|r| r|r|r| r|r|r| r|r|r|}
			\hline 	\multicolumn{3}{|c|}{\rm{Model~\hyperref[model6b]{5b}}}  & \multicolumn{11}{c|}{{\rm U}(4)\times {\rm USp}(2)_L \times {\rm U}(2)_R \times {\rm U}(2)\times {\rm USp}(2)}\\
			\hline \hline \rm{stack} & N & (n^1,l^1)\times (n^2,l^2)\times (n^3,l^3) & n_{\yng(2)} & n_{\yng(1,1)_{}} & b & c & c'& d & d' & 1 & 2 & 3 & 4 \\
			\hline
			a     & 8     & (1, 3)\times (1, 0)\times (1, -1) & -2    & 2       & 3     & 0     & -3    & 4     & 4     & 0     & 0     & 0     & 0 \\
			b     & 2     & (1, 0)\times (0, 1)\times (0, -2) & 0     & 0       & \text{-}    & -1    & 1     & -1     & 1    & 0     & 0     & 0     & 0 \\
			c     & 4     & (0, -1)\times (-1, 3)\times (1, -1) & 2     & -2       & \text{-}    & \text{-}    & \text{-}    & -4     & 4   & -3     & 0     & 0     & 0 \\
			d     & 4     & (1, 1)\times (-1, 1)\times (-1, -3) & 0    & 12      & \text{-}    & \text{-}    & \text{-}    & \text{-}    & \text{-}    & 3     & 0     & 0     & 0 \\
			\hline
			1     & 2     & (1, 0)\times (1, 0)\times (2, 0) &  \multicolumn{11}{c|}{\beta^g_{d}=7,~ \beta^g_1=0,  } \\
			&       &                                   &   \multicolumn{11}{c|}{x_A=\frac{1}{3}x_B=\frac{1}{33}x_C=\frac{1}{9}x_D }  \\
			&       &                                   &   \multicolumn{11}{c|}{33\chi_1=3\chi_2=\frac{11}{2}\chi_3= \sqrt{11} }  \\
			\hline
		\end{array}$
	\end{table}
\end{widetext}


\end{document}